\documentclass{JHEP3}
\input epsf
\usepackage{epsfig}
\usepackage{amssymb}
\usepackage{axodraw}
\usepackage[active]{srcltx}

\newcommand{\bea}{\begin{eqnarray}}
\newcommand{\eea}{\end{eqnarray}}
\newcommand{\bean}{\begin{eqnarray*}}
\newcommand{\eean}{\end{eqnarray*}}
\newcommand{\nn}{\nonumber \\}

\def\W #1{\widetilde{#1}}

\def\ket#1{\left| #1\right\rangle}
\def\gb #1{ \left\langle #1 \right]}
\def\tgb #1{ \left[ #1 \right\rangle}
\def\cb #1{ \left[ #1 \right]}
\def\trm{\mathop{{\rm tr}_-}}

\def\Tr{\mathop{\rm Tr}}

\def\a{{\alpha}}

\def\b{{\beta}}

\def\eps{\epsilon}

\def\vev#1{\left\langle #1 \right\rangle}

\def\Label#1{\label{#1}}

\def\spa#1.#2{\langle#1\,#2\rangle}
\def\spb#1.#2{[#1\,#2]}
\def\spab#1.#2.#3{\langle\mskip-1mu{#1}
                  | #2 | {#3}]}

\def\spba#1.#2.#3{[\mskip-1mu{#1}
                  | #2 | {#3}\rangle}

\def\spbb#1.#2.#3.#4{[\mskip-1mu{#1}
                     | {#2} \ {#3} | {#4}]}

\def\spaa#1.#2.#3.#4{\langle\mskip-1mu{#1}
                     | {#2} \ {#3} | {#4}\rangle}

\newbox\SlashedBox
\def\slashed#1{\setbox\SlashedBox=\hbox{#1}
\hbox to 0pt{\hbox to 1\wd\SlashedBox{\hfil/\hfil}\hss}#1}
\def\hboxtosizeof#1#2{\setbox\SlashedBox=\hbox{#1}
\hbox to 1\wd\SlashedBox{#2}}

\newbox\charbox
\newbox\slabox
\def\s#1{{      
        \setbox\charbox=\hbox{$#1$}
        \setbox\slabox=\hbox{$/$}
        \dimen\charbox=\ht\slabox
        \advance\dimen\charbox by -\dp\slabox
        \advance\dimen\charbox by -\ht\charbox
        \advance\dimen\charbox by \dp\charbox
        \divide\dimen\charbox by 2
        \raise-\dimen\charbox\hbox to \wd\charbox{\hss/\hss}
        \llap{$#1$}
}}

\preprint{CERN-PH-TH/2008-050\\ITFA-2008-06}
\title{Closed-Form Decomposition of One-Loop Massive Amplitudes}
\author{Ruth Britto$^{a}$, Bo Feng$^{b}$, Pierpaolo Mastrolia$^{c}$\\
~~~~\\
$^a$Institute for Theoretical Physics, University of Amsterdam \\
Valckenierstraat 65, 1018 XE Amsterdam, The Netherlands\\
$^b$Center of Mathematical Science, Zhejiang University, Hangzhou, China\\
$^c$Theory Division, CERN CH-1211 Geneva 23, Switzerland
}
\abstract{
We present formulas for the coefficients of 2-, 3-, 4- and 5-point master integrals for one-loop massive amplitudes.
The coefficients are derived from unitarity cuts in $D$ dimensions.
The input parameters 
can be read off from any unitarity-cut integrand, as
assembled from tree-level expressions, after
simple algebraic manipulations.
The formulas presented here are suitable
for analytical as well as  numerical
evaluation.
Their validity is confirmed in two
known cases of helicity amplitudes contributing
to $gg \to gg$, $gg \to gH$, where the masses of the Higgs and
 the fermion circulating in the loop
 are kept as free parameters.
}
\keywords{NLO Computations, QCD}

\begin{document}

\section{Introduction}

The unitarity method introduced in~\cite{Bern:1994zx, Bern:1994cg} is designed to
compute any scattering amplitude by matching its unitarity cuts
onto the corresponding cuts of
its expansion in a basis of master integrals~\cite{MasterIntegrals}
with rational coefficients.  Each of these coefficients can be
determined quantitatively from prior knowledge of the master integrals and the singularity structure of the amplitude.

As the master integrals form a basis for amplitudes, so
the unitarity cuts of master integrals
have uniquely identifiable analytic properties, and can be
used as a basis for the cuts of any amplitude.
Therefore, the coefficients of the linear combination can be extracted
systematically through the
phase-space integration (instead of
complete loop integration).

Recently, unitarity-based methods for one-loop
amplitudes have been the subject of an intense investigation,
through different implementations of the cut-constraints
\cite{Cachazo:2004by,Bena:2004xu,Cachazo:2004dr,Britto:2004nj,Britto:2004nc,Britto:2005ha,Brandhuber:2005jw,Britto:2006sj,Anastasiou:2006jv,Mastrolia:2006ki,Britto:2006fc,Anastasiou:2006gt,Britto:2007tt,OPP,Forde:2007mi,Kilgore:2007qr,BjerrumBohr:2007vu,Ellis:2007br,Giele:2008ve}.

The holomorphic anomaly of
 unitarity cuts~\cite{Cachazo:2004by,Cachazo:2004kj}
simplifies the phase-space integration dramatically: cut-integrals
can be done analytically by evaluating residues of a complex
function in spinor variables~\cite{SpinorFormalism},  reducing the problem of so-called tensor-reduction to one
of {\em algebraic} manipulation.

Accordingly, in~\cite{Britto:2005ha,Britto:2006sj},
a systematic method was introduced to evaluate any finite
four-dimensional unitarity cut, yielding  compact
expressions for the coefficients of the master integrals.
This method was successfully applied to the final parts of the cut-constructible part of the six-gluon amplitude in QCD.
The same method, based on the spinor-integration
of the phase-space, was later extended for the evaluation of
generalised cuts in $D$ dimensions~\cite{Anastasiou:2006jv,Mastrolia:2006ki,Britto:2006fc,Anastasiou:2006gt},
which is essential for the complete determination
of any amplitude in dimensional regularization
\cite{DDimU,Bern:1995db,Rozowsky:1997dm}.

In this paper, we carry out the extension
to the massive case of the analytic results presented in \cite{Britto:2007tt},
stemming from an original study
of compact formulas for the coefficients of the master integrals
\cite{Britto:2006fc}.
Following the same logic as in \cite{Britto:2007tt},
we now present general formulas for the coefficients of
the master integrals which can be evaluated
without performing any integration.
These formulas depend on input variables (indices, momenta and associated spinors)
that are specific to the initial cut-integrand, which is
assembled from tree-level amplitudes.
The value of a given coefficient is thus obtained simply
by pattern-matching, that is
by specializing the value of the input variables to be
inserted in the general formulas. 
The implementation of the general formulas into automatic tools
is straightforward, as
done for the current investigation with the program
{\tt S@M}~\cite{Maitre:2007jq}.

In this paper, since the formulas for the
coefficients are obtained via massive double cuts in $D$-dimension,
we do not present results for the coefficients of
cut-free functions like
tadpoles and bubbles with massless external
momentum (which can be expressed in terms of tadpoles as well).
The coefficients of such functions could be fixed either
by imposing the expected UV-behaviour of the amplitude, as
described in \cite{Bern:1995db},
or computed with other techniques applicable in
massive calculations~\cite{OPP,Forde:2007mi,Kilgore:2007qr,Ellis:2007br,Giele:2008ve}.

The paper is organized as follows. In section 2, we describe the
structure of the decomposition of one-loop amplitude in terms of
master integrals. In Section 3 we explain the double-cut integration
with spinor variables, which leads to the formulas of the
coefficients of the master integrals, presented in Section 4. In
Sections 5 and 6, we apply our formulas to two examples of one-loop
scattering amplitudes, respectively $gH \to gg$ and $gg \to gg$,
where the Higgs mass and the mass of the internal fermion (in both
cases) are kept as free parameters. 
In Section 7, we present both analytical and numerical methods 
to obtain, finally, the explicit coefficients of the dimensionally
shifted master integrals.
In Appendix A, we record the
translation between our basis of integrals and the ones
used in the literature for the examples discussed in  Sections 5 and 6.
In Appendix B, we present a proof of the decomposition into the
dimensionally shifted basis, with rational coefficients independent of
 $\eps$. In other words, we prove that the coefficients given
by our algebraic expressions will be polynomial in our
extra-dimensional variable $u$. As
a byproduct, we have produced equivalent and simpler algebraic
functions for the evaluation of coefficients. 

\section{Decomposition in terms of master integrals}

We define the
 $n$-point scalar function with non-uniform masses as follows:\footnote{For ease of presentation, we are omitting the prefactor $i(-1)^{n+1}(4\pi)^{D/2}$ (which was included for example in \cite{Bern:1995db}).}
\bea   I_n(M_1,M_2,m_1,\ldots,m_{n-2})  \equiv
 \int {d^{4-2\eps} p \over
(2\pi)^{4-2\eps}}{1\over (p^2-M_1^2) ((p-K)^2-M_2^2)
\prod_{j=1}^{n-2} ((p-P_j)^2-m_{j}^2)}.~~~\label{n-scalar} \eea
Giele, Kunszt and Melnikov \cite{Giele:2008ve} have given the decomposition of
any one-loop amplitude in $D$ dimensions in terms of master integrals,
represented here pictorially.

\bea
\begin{picture}(0,0)(0,0)
\SetScale{0.65}
\SetWidth{1.0}
\GOval(0,0)(27,27)(0){1}
\Line(-20,-20)(-30,-30)
\Line(20,-20)(30,-30)
\Line(20,20)(30,30)
\Line(-20,20)(-30,30)
\Text(0,0)[]{{\tiny{$A_n^{(D)}$}}}
\end{picture}
\hspace*{1.0cm}
&=&
\qquad \!
e^{(0)} \hspace*{1cm}
\begin{picture}(0,0)(0,0)
\SetScale{0.65}
\SetWidth{1.0}
\Line(-17,-20)(17,-20)
\Line(-17,-20)(-25,10)
\Line(17,-20)(25,10)
\Line(-25,10)(0,25)
\Line(+25,10)(0,25)
\Line(0,25)(0,35)
\Line(-25,10)(-32,15)
\Line(+25,10)(+32,15)
\Line(-17,-20)(-25,-30)
\Line( 17,-20)(25,-30)
\Text(0,0)[]{{\tiny{$I_5^{(D)}$}}}
\end{picture}
\nonumber \\ & &
\nonumber \\ & &
+ \quad
d^{(0)} \hspace*{1cm}
\begin{picture}(0,0)(0,0)
\SetScale{0.65}
\SetWidth{1.0}
\Line(-20,-20)(20,-20)
\Line(20,-20)(20,20)
\Line(20,20)(-20,20)
\Line(-20,20)(-20,-20)
\Line(-20,-20)(-30,-30)
\Line(20,-20)(30,-30)
\Line(20,20)(30,30)
\Line(-20,20)(-30,30)
\Text(0,0)[]{{\tiny{$I_4^{(D)}$}}}
\end{picture}
\hspace*{1cm} +
\quad
d^{(2)} \hspace*{1cm}
\begin{picture}(0,0)(0,0)
\SetScale{0.65}
\SetWidth{1.0}
\Line(-20,-20)(20,-20)
\Line(20,-20)(20,20)
\Line(20,20)(-20,20)
\Line(-20,20)(-20,-20)
\Line(-20,-20)(-30,-30)
\Line(20,-20)(30,-30)
\Line(20,20)(30,30)
\Line(-20,20)(-30,30)
\Text(0,0)[]{{\tiny{$I_4^{(D+2)}$}}}
\end{picture}
\hspace*{1cm} +
\quad
d^{(4)} \hspace*{1cm}
\begin{picture}(0,0)(0,0)
\SetScale{0.65}
\SetWidth{1.0}
\Line(-20,-20)(20,-20)
\Line(20,-20)(20,20)
\Line(20,20)(-20,20)
\Line(-20,20)(-20,-20)
\Line(-20,-20)(-30,-30)
\Line(20,-20)(30,-30)
\Line(20,20)(30,30)
\Line(-20,20)(-30,30)
\Text(0,0)[]{{\tiny{$I_4^{(D+4)}$}}}
\end{picture}
\nonumber \\ & &
\nonumber \\ & &
 +
\quad
c^{(0)}\hspace*{1cm}
\begin{picture}(0,0)(0,0)
\SetScale{0.65}
\SetWidth{1.0}
\Line(-20,-20)(20,0)
\Line(-20,20)(20,0)
\Line(-20,-20)(-20,20)
\Line(-20,-20)(-30,-30)
\Line(20,0)(30,10)
\Line(20,0)(30,-10)
\Text(-5,0)[]{{\tiny{$I_3^{(D)}$}}}
\Line(-20,20)(-30,30)
\end{picture}
\hspace*{1cm} +
\quad
c^{(2)}\hspace*{1cm}
\begin{picture}(0,0)(0,0)
\SetScale{0.65}
\SetWidth{1.0}
\Line(-20,-20)(20,0)
\Line(-20,20)(20,0)
\Line(-20,-20)(-20,20)
\Line(-20,-20)(-30,-30)
\Line(20,0)(30,10)
\Line(20,0)(30,-10)
\Text(-3,0)[]{{\tiny{$I_3^{\!(\!D\!+\!2\!)}$}}}
\Line(-20,20)(-30,30)
\end{picture}
\nonumber \\ & &
\nonumber \\ & &
 +
\quad
b^{(0)}\hspace*{1.0cm}
\begin{picture}(0,0)(0,0)
\SetScale{0.65}
\SetWidth{1.0}
\Line(-20,0)(-30,10)
\Line(-20,0)(-30,-10)
\Line(20,0)(30,10)
\Line(20,0)(30,-10)
\Text(0,0)[]{{\tiny{$I_2^{(D)}$}}}
\Oval(0,0)(20,20)(0)
\end{picture}
\hspace*{1.0cm} +
\quad
b^{(2)}\hspace*{1.0cm}
\begin{picture}(0,0)(0,0)
\SetScale{0.65}
\SetWidth{1.0}
\Line(-20,0)(-30,10)
\Line(-20,0)(-30,-10)
\Line(20,0)(30,10)
\Line(20,0)(30,-10)
\Text(0,0)[]{{\tiny{$I_2^{(D+2)}$}}}
\Oval(0,0)(20,20)(0)
\end{picture}
\hspace*{1.0cm} + \quad
a^{(0)}\hspace*{1.0cm}
\begin{picture}(0,0)(0,0)
\SetScale{0.65}
\SetWidth{1.0}
\Line(0,-20)(-10,-30)
\Line(0,-20)( 10,-30)
\Text(0,0)[]{{\tiny{$I_1^{(D)}$}}}
\Oval(0,0)(20,20)(0)
\end{picture}
%
\label{DdimMasterDeco}
\eea

\vspace*{0.5cm}

\noindent
Here,
with reference to \cite{Giele:2008ve}:
{\it i)} we have absorbed the residual $D$-dependence of the coefficients
in the definition of the master integrals;
{\it ii)} for ease of notation, we have given as understood the sums on the partition
of the $n$-points of the amplitude in the number of points
corresponding to each master integral.
Thus, the coefficients $e,d,c,b,a$ in Eq.(\ref{DdimMasterDeco}) are independent
of $D$.

If, on both sides of Eq.(\ref{DdimMasterDeco}), we apply the
standard decomposition of the $D=4-2\eps$ dimensional loop variable, $L$,
in a four-dimensional
component, $\tilde{\ell}$, and its $(-2\eps)$-dimensional orthogonal complement, $\mu$,
\bea
L = \W \ell + \mu \ .
\Label{L-stddeco}
\eea
then the integration measure becomes
\bea
\int d^{4-2\eps} L = \int d^{-2\eps}\mu  \ \int d^4 \W \ell \ ,
\eea
namely the composition of a four-dimensional integration and
an integration over a $(-2\eps)$-dimensional mass-like parameter.
By taking the $\mu$-integral to be understood,
the four-dimensional integration on both sides of Eq.(\ref{DdimMasterDeco}),
can be read as follows:

\bea
\begin{picture}(0,0)(0,0)
\SetScale{0.65}
\SetWidth{1.0}
\GOval(0,0)(27,27)(0){1}
\Line(-20,-20)(-30,-30)
\Line(20,-20)(30,-30)
\Line(20,20)(30,30)
\Line(-20,20)(-30,30)
\Text(0,0)[]{{\tiny{$A_n^{(4)}$}}}
\end{picture}
\hspace*{1.0cm}
&=&
\Big(e^{(0)} \ \pi(\mu^2) + d_2(\mu^2) \Big) \hspace*{0.7cm}
\begin{picture}(0,0)(0,0)
\SetScale{0.65}
\SetWidth{1.0}
\Line(-20,-20)(20,-20)
\Line(20,-20)(20,20)
\Line(20,20)(-20,20)
\Line(-20,20)(-20,-20)
\Line(-20,-20)(-30,-30)
\Line(20,-20)(30,-30)
\Line(20,20)(30,30)
\Line(-20,20)(-30,30)
\Text(0,0)[]{{\tiny{$I_4^{(4)}$}}}
\end{picture}
\hspace*{1.0cm} +
c_1(\mu^2)\hspace*{0.7cm}
\begin{picture}(0,0)(0,0)
\SetScale{0.65}
\SetWidth{1.0}
\Line(-20,-20)(20,0)
\Line(-20,20)(20,0)
\Line(-20,-20)(-20,20)
\Line(-20,-20)(-30,-30)
\Line(20,0)(30,10)
\Line(20,0)(30,-10)
\Text(-5,0)[]{{\tiny{$I_3^{(4)}$}}}
\Line(-20,20)(-30,30)
\end{picture}
\hspace*{1.0cm} +
b_1(\mu^2)\hspace*{1.0cm}
\begin{picture}(0,0)(0,0)
\SetScale{0.65}
\SetWidth{1.0}
\Line(-20,0)(-30,10)
\Line(-20,0)(-30,-10)
\Line(20,0)(30,10)
\Line(20,0)(30,-10)
\Text(0,0)[]{{\tiny{$I_2^{(4)}$}}}
\Oval(0,0)(20,20)(0)
\end{picture}
\hspace*{1.0cm} +
a^{(0)}\hspace*{0.7cm}
\begin{picture}(0,0)(0,0)
\SetScale{0.65}
\SetWidth{1.0}
\Line(0,-20)(-10,-30)
\Line(0,-20)( 10,-30)
\Text(0,0)[]{{\tiny{$I_1^{(4)}$}}}
\Oval(0,0)(20,20)(0)
\end{picture}
%
\label{4dimMasterDeco}
\eea

\vspace*{0.5cm}

\noindent
where
$d_n(\mu^2), c_n(\mu^2),$ and $b_n(\mu^2)$ are polynomials of
degree $n$ in $\mu^2$, as discussed in  Appendix
\ref{polynomialproof}, 
\bea
d_2(\mu^2) &=& d^{(0)} + d^{(2)} \mu^2 + d^{(4)} (\mu^2)^2 \ , \\
c_1(\mu^2) &=& c^{(0)} + c^{(2)} \mu^2  \ , \\
b_1(\mu^2) &=& b^{(0)} + b^{(2)} \mu^2  \ ,
\eea
whereas $\pi(\mu^2)$ is non-polynomial in $\mu^2$ and  corresponds to
the coefficients of the reduction of the pentagon to boxes, which
occurs in $D=4$. 

The polynomial structure of $d_n(\mu^2), c_n(\mu^2),$ and $b_n(\mu^2)$
is responsible for the dimensionally shifted integrals appearing in
Eq.(\ref{DdimMasterDeco}), because the $\mu$-integration can be performed
trivially by absorbing the extra powers of $\mu^2$ into the integration measure, according to
\cite{Bern:1995db}:
\bea
\int {d^{ - 2 \eps} \mu \over (2\pi)^{ - 2 \eps}} \ (\mu^2)^r \ f( \mu^2)
&=&
- \eps (1-\eps) (2-\eps) \cdots (r-1-\eps) (4\pi)^r
\int {d^{2r - 2 \eps} \mu \over (2\pi)^{2r - 2 \eps}} \ f( \mu^2)  \ .
\eea
The presence of $\pi(\mu^2)$ in the coefficient of the four-dimensional box 
is a unique signature of the pentagon.
We conclude that the reconstruction of the
four-dimensional kernel of any one-loop amplitude, given in Eq.(\ref{4dimMasterDeco}),
contains all the information for the complete
reconstruction of the amplitude in $D$-dimensions, given in Eq.(\ref{4dimMasterDeco}).

In the following pages, we present the general formulas of the
coefficients of the
box, $I_4^{(4)}$, triangle, $I_3^{(4)}$, and bubble, $I_2^{(4)}$,
obtained from the double cut of
Eq.(\ref{4dimMasterDeco}), 

\bea
\begin{picture}(0,0)(0,0)
\SetScale{0.65}
\SetWidth{1.0}
\GOval(0,0)(27,27)(0){1}
\Line(-20,-20)(-30,-30)
\Line(20,-20)(30,-30)
\Line(20,20)(30,30)
\Line(-20,20)(-30,30)
\Text(0,0)[]{{\tiny{$A_n^{(4)}$}}}
\DashLine(0,30)(0,-30){3}
\end{picture}
\hspace*{1.0cm}
&=&
\Big(e^{(0)} \ \pi(\mu^2) + d_2(\mu^2) \Big) \hspace*{0.7cm}
\begin{picture}(0,0)(0,0)
\SetScale{0.65}
\SetWidth{1.0}
\Line(-20,-20)(20,-20)
\Line(20,-20)(20,20)
\Line(20,20)(-20,20)
\Line(-20,20)(-20,-20)
\Line(-20,-20)(-30,-30)
\Line(20,-20)(30,-30)
\Line(20,20)(30,30)
\Line(-20,20)(-30,30)
\Text(0,0)[]{{\tiny{$I_4^{(4)}$}}}
\DashLine(0,30)(0,-30){3}
\end{picture}
\hspace*{1.0cm} +
c_1(\mu^2)\hspace*{0.7cm}
\begin{picture}(0,0)(0,0)
\SetScale{0.65}
\SetWidth{1.0}
\Line(-20,-20)(20,0)
\Line(-20,20)(20,0)
\Line(-20,-20)(-20,20)
\Line(-20,-20)(-30,-30)
\Line(20,0)(30,10)
\Line(20,0)(30,-10)
\Text(-5,0)[]{{\tiny{$I_3^{(4)}$}}}
\Line(-20,20)(-30,30)
\DashLine(0,30)(0,-30){3}
\end{picture}
\hspace*{1.0cm} +
b_1(\mu^2)\hspace*{1.0cm}
\begin{picture}(0,0)(0,0)
\SetScale{0.65}
\SetWidth{1.0}
\Line(-20,0)(-30,10)
\Line(-20,0)(-30,-10)
\Line(20,0)(30,10)
\Line(20,0)(30,-10)
\Text(0,0)[]{{\tiny{$I_2^{(4)}$}}}
\Oval(0,0)(20,20)(0)
\DashLine(0,30)(0,-30){3}
\end{picture}
%
\label{4dimCutMasterDeco}
\eea

\vspace*{0.5cm}

\noindent
Since the formulas for the
coefficients are obtained via double cuts,
we do not present the results for the coefficients of
cut-free functions like
tadpoles and bubbles with massless external
momentum (which can be expressed in terms of tadpoles as well).
Their coefficients could be fixed either
by imposing the expected UV-behaviour of the amplitude, as
described in \cite{Bern:1995db},
or computed with alternative techniques~\cite{OPP,Forde:2007mi,Kilgore:2007qr,Ellis:2007br,Giele:2008ve}.

\section{The double cut phase space integration}

In this section, we review the $D$-dimensional unitarity method
\cite{Anastasiou:2006jv, Anastasiou:2006gt} as applied in cases with
arbitrary masses \cite{Britto:2006fc}.  Our goal is to describe the
structure of the cut integrand, from which we will directly read off
the coefficients from the formulas in the following section.  The
formulas will be the massive analogs of the ones in \cite{Britto:2007tt}.

Recall the phase space integration of a standard (double) cut in $D=4-2\eps$
dimensions.
We use the usual decomposition of
the $D$-dimensional loop variable, $L$,
in a four-dimensional component, $\W \ell$, and
a transverse $(-2\eps)$-dimensional remnant, $\mu$,
\bea
L = \W \ell + \mu \ .
\Label{L-stddeco-2}
\eea
The integration measure becomes
\bea
\int d^{4-2\eps} L = \int d^{-2\eps}\mu \int d^4 \W \ell
= {(4\pi)^{\eps} \over \Gamma(-\eps)}
\int d\mu^2 \ (\mu^2)^{-1-\eps} \int d^4 \W \ell \ ,
\eea
namely the composition of a four-dimensional integration and
an integration over a $(-2\eps)$-dimensional mass-like parameter.
In order to write the four-dimensional part in terms of spinor variables
associated to massless momentum,
we proceed with the following change of variables:
\bea \W \ell= \ell+z K,~~~~~\ell^2=0,
~~~\Label{changing} \eea
where $\ell$ is a massless momentum and
$K$ is the momentum across the cut, fixed by the kinematics.
Accordingly, the four-dimensional integral measure becomes
\bea
\int d^4\W \ell= \int dz~ d^4\ell~ \delta^+(\ell^2) (2 \ell \cdot K) \ .
\eea
The Lorentz-invariant phase-space (LIPS) of a double cut in the
$K^2$-channel is defined by
the presence of two $\delta$-functions imposing the cut conditions:
\bea
\int d^{4-2\eps} \Phi =
\int d^{4-2\eps} L \
\delta(L^2-M_1^2) \ \delta((L-K)-M_2^2).
\eea
Here $M_1$ and $M_2$ are the masses of the cut lines.
By using the decomposition of the loop variable
in Eq.(\ref{L-stddeco}), the four-dimensional integral can be
separated, so that
\bea
\int d^{4-2\eps} \Phi
= {(4\pi)^{\eps} \over \Gamma(-\eps)}
\int d\mu^2 \ (\mu^2)^{-1-\eps} \
\int d^{4} \phi,
\eea
where the four-dimensional LIPS is
\bea
\int d^{4} \phi & = &  \int d^4 \W \ell \
\delta(\W \ell^2- M_1^2-\mu^2) \ \delta((\W \ell-K)^2-M_2^2-\mu^2) \ .
\eea
The change of variables in Eq.(\ref{changing}),
and the $z$-integration (trivialized by the presence
of $\delta$'s),
yield the four-dimensional LIPS to appear as
\bea
\int d^4 \phi &=&
\int  d^4\ell \
\delta^+(\ell^2) \
\delta((1-2z)K^2-2 \ell \cdot K+M_1^2-M_2^2),
\Label{double-cut-measure}
\eea
where
\bea
z = { (K^2+M_1^2-M_2^2)- \sqrt{\Delta[K_1, M_1, M_2]- 4 K^2
\mu^2}\over 2 K^2},~~~~\Label{solve-z}
\eea
with
\bea \Delta[K,M_1,M_2]\equiv (K^2)^2+(M_1^2)^2+(M_2^2)^2-2 K^2 M_1^2
-2 K^2 M_2^2- 2M_1^2 M_2^2 \ .
~~~\Label{Delta-KMM}
\eea
We remark that the value of $z$ in Eq.(\ref{solve-z})
is frozen to be
the proper root ($K>0$) of the quadratic argument of
$\delta(z(1-z)K^2 + z(M_1^2-M_2^2)-M_1^2-\mu^2)$, coming from
$\delta(\W \ell^2- M_1^2-\mu^2)$.
For later convenience, one can redefine the $\mu^2$-integral measure as
\bean \int d\mu^2 (\mu^2)^{-1-\eps} = \left( {
\Delta[K,M_1,M_2]\over 4 K^2}\right)^{-\eps}
\int_0^1 du \ u^{-1-\eps},
\eean
where the relation between $u$ and $\mu^2$ is given by 
\bea u\equiv {4 K^2\mu^2 \over
\Delta[K,M_1,M_2]},~~~~~~~\mu^2=\left( { u\Delta[K,M_1,M_2]\over 4
K^2}\right).~~~\Label{u-def}\eea
We observe that the domain of $u$, i.e., $u\in [0,1]$,
follows from the kinematical constraints, as discussed in \cite{Britto:2006fc}.

Finally, after the above rearrangement, the $D$-dimensional
Lorentz-invariant phase-space of a double cut in the
$K^2$-channel can be written in a suitable form,
\bea
\int d^{4-2\eps} \Phi
&=&
\chi(\eps,K,M_1,M_2) \
\int_0^1 du \ u^{-1-\eps}
\int d^4\phi,
\label{Ddim:2PLEcut}
\eea
where
\bea
\chi(\eps,K,M_1,M_2) &=&
{(4\pi)^{\eps} \over \Gamma(-\eps)}
\left( {\Delta[K,M_1,M_2]\over 4 K^2} \right)^{-\eps},
\eea
and where $d^4\phi$ was given in Eq.(\ref{double-cut-measure}).
By using the definition of $u$ given in Eq.(\ref{u-def}), we can write
\bea z ={ \a -\b \sqrt{1-u}\over 2},~~~~\Label{z-sol-u}\eea
where
\bea \a = { K^2+M_1^2-M_2^2\over K^2},~~~~\b={\sqrt{\Delta[K, M_1,
M_2]}\over K^2}.~~~\Label{ab-def}\eea
Notice that when $M_1=M_2=0$ we have $\a=\b=1$, thus
reproducing the massless case.
A useful relation between $z$ and $u$ is the following:
\bea (1-2z) +{ M_1^2-M_2^2\over K^2} = \b
\sqrt{1-u}.~~~\Label{z-rel-1}\eea
This relation will be used in  Appendix \ref{polynomialproof} to prove that
that the coefficients given in this paper are
polynomials in $u$, or equivalently $\mu^2$.  As discussed in the
previous section,
this feature is essential for the straightforward
reconstruction of dimensionally-shifted master integrals.

\bigskip


The main feature of a double-cut LIPS parametrized as in
Eqs(\ref{Ddim:2PLEcut},\ref{double-cut-measure}) is that
the kernel of the integration is represented by
the four-dimensional integral.
In fact, the $u$-integration (or equivalently, the $\mu^2$-integration),
is simply responsible of the rise of {\it shifted-dimension} master integrals.
Thus, our interest in
the extraction of the coefficients of the master integrals
from a four-dimensional massive double cut, see Eq.(\ref{4dimCutMasterDeco}),
translates in focusing the discussion only on the $\int d^4\phi$.

The $D$-dimensional double cut of any one-loop amplitude
is, in general form,
\bea
\int d^{4-2\eps} \Phi \ A_L^{\rm tree} \times A_R^{\rm tree}
=
\chi(\eps,K,M_1,M_2) \
\int_0^1 du \ u^{-1-\eps}
\int d^4\phi
\ A_L^{\rm tree} \times A_R^{\rm tree}
\eea
where
$A_L^{\rm tree}$ and $A_R^{\rm tree}$ are the two tree-level amplitudes 
on the left and right side of the cut.
As discussed above,
the kernel of the integration is represented by the four-dimensional
part,
\bea
\int d^4\phi
\ A_L^{\rm tree} \times A_R^{\rm tree} \ .
\eea
We proceed from the formula (\ref{double-cut-measure}) by
introducing spinor-variables according to \cite{Cachazo:2004kj},
\bea
\int d^4\ell \ \delta(\ell^2) &=& \int \vev{\ell~d\ell}[\ell~d\ell]
\int t ~dt
\eea
and performing the integral over $t$ trivially, with the second delta function.
The general expression of
the double cut integral will then be
\bea
& & \int d^4\phi
\ A_L^{\rm tree} \times A_R^{\rm tree}
\nonumber
\\
&=& \int  d^4\ell \delta^+(\ell^2)\delta((1-2z)K^2-2 \ell
\cdot K+M_1^2-M_2^2) { \prod_j \gb{a_j|\W\ell|b_j}\over \prod_i ((\W\ell-K_i)^2-m_i^2-\mu^2)}
~~~~\Label{gen-integrand}
\\
& = & \int  d^4\ell \delta^+(\ell^2)\delta((1-2z)K^2-2 \ell \cdot
K+M_1^2-M_2^2) { \prod_j \gb{a_j|\ell+zK|b_j}\over \prod_i (
K_i^2+M_1^2-m_i^2-2 (\ell+zK)\cdot K_i)}
\nonumber
\\& = & \int
\vev{\ell~d\ell}[\ell~d\ell] \int t dt
\delta((1-2z)K^2+t\gb{\ell|K|\ell}+M_1^2-M_2^2) { \prod_j (z\gb{a_j|
K|b_j}+ t\gb{\ell|P_j|\ell}) \over \prod_i ( K_i^2+M_1^2-m_i^2-2 z
K\cdot K_i+t \gb{\ell|K_i|\ell})} \nonumber
\eea
Here we have used $\W\ell^2= M_1^2+\mu^2$. Notice
that $\gb{a|\W\ell|b}=-2\W\ell\cdot P$, with $P=|a\rangle [b|$.

After using the remaining delta function to perform the integral over $t$, we have
\bea \int \vev{\ell~d\ell}[\ell~d\ell]\left((1-2z)+{M_1^2-M_2^2\over
K^2} \right){ (K^2)^{n+1}\over \gb{\ell|K|\ell}^{n+2}}
{\prod_{j=1}^{n+k} \gb{\ell|R_j|\ell}\over \prod_{i=1}^k
\gb{\ell|Q_i|\ell}}~~~\Label{frame}\eea
where
\bea
P_j&=&|a_j\rangle [b_j| \ , \\
R_j & = &-\left((1-2z)+{M_1^2-M_2^2\over K^2} \right)P_j- {
z(2P_j\cdot K)\over
K^2}K,~~~~\Label{R-def} \\
Q_j & = & -\left((1-2z) +{M_1^2-M_2^2\over K^2} \right)K_j+
{K_j^2+M_1^2-m_j^2-2 z K\cdot K_j\over K^2}K~~~~\Label{Q-def}
\eea
The vectors $P_j, R_j, Q_j$ do not depend on the loop variables, but rather only on
the external kinematics, and especially on the momentum across the cut, $K$.
Applying (\ref{frame}) to the master integrals, we find the following results.
\begin{itemize}

\item (1) Bubble:  $k=0$, $n+k=0$.
\bea \int \vev{\ell~d\ell}[\ell~d\ell]\left((1-2z)+{M_1^2-M_2^2\over
K^2} \right){ (K^2)\over \gb{\ell|K|\ell}^2}\eea

\item (2) Triangle:  $k=1$, $n+k=0$.
\bea \int \vev{\ell~d\ell}[\ell~d\ell]\left((1-2z)+{M_1^2-M_2^2\over
K^2} \right){ 1\over \gb{\ell|K|\ell}\gb{\ell|Q|\ell}}\eea

\item (3) Box: $k=2$, $n+k=0$.
\bea \int \vev{\ell~d\ell}[\ell~d\ell]\left((1-2z)+{M_1^2-M_2^2\over
K^2} \right){ (K^2)^{-1}\over
\gb{\ell|Q_1|\ell}\gb{\ell|Q_2|\ell}}\eea

\end{itemize}

These formulas are the extension to the massive case of the
corresponding ones given in \cite{Britto:2007tt}. We notice that the
presence of the masses enters only the definitions of $P_j$, $R_j$,
and $Q_j$. Therefore the spinor integration performed in the
massless case \cite{Britto:2007tt} is valid as well in this case.

\bigskip

The expression of the cut-integrand in Eq.(\ref{frame}),
with its indices, $n$ and $k$, and its vectors $P_j$, $R_j$,
and $Q_j$
is the key to constructing the coefficients.
In the next section we present
general formulas for the coefficients of the master-integrals
(boxes, triangles and bubbles),
which depend on exactly these input parameters.
Accordingly,
given a specific amplitude (or integral), one can
obtain its decomposition in terms of master-integrals
{\em without any integration}.
Every coefficient is obtained from
the general formulas simply by substituting
 the input parameters characterizing
the specific amplitude.  These parameters  are obtained by pattern-matching
onto the reference form in Eq.(\ref{frame}).


\section{\label{sec:formulas}Formulas for the coefficients of master integrals}

The coefficients of master integrals are obtained by the procedure described in the previous section, which is a straightforward generalization of the massless case \cite{Britto:2007tt}.
We list the results in this section.
In fact, the expressions take the same form as in the massless
case; the mass dependence enters directly through the definitions (\ref{R-def})
and (\ref{Q-def}), and through these formulas into the definitions
(\ref{box-null}), (\ref{tri-null}).

\subsection{Box coefficient}

\vspace*{1cm}

\begin{figure}[h]
\begin{center}
\begin{picture}(0,0)(0,0)
\SetScale{0.7}
\SetWidth{1.0}
\Line(-20,-20)(20,-20)
\Line(20,-20)(20,20)
\Line(20,20)(-20,20)
\Line(-20,20)(-20,-20)
\Line(-20,-20)(-30,-30)
\Line(20,-20)(30,-30)
\Line(20,20)(30,30)
\Line(-20,20)(-30,30)
\DashLine(0,30)(0,-30){3}
\Text(30,30)[]{\footnotesize{$K_r$} }
\Text(-30,+30)[]{\footnotesize{$K_s$} }
\Text(+40,-30)[]{\footnotesize{$K-K_r$} }
\end{picture}
\hspace*{4.0cm}
\begin{picture}(0,0)(0,0)
\SetScale{0.7}
\SetWidth{1.0}
\Line(-25,0)(0,25)
\Line(0,25)(25,0)
\Line(25,0)(0,-25)
\Line(0,-25)(-25,0)
\Line(-35,0)(-25,0)
\Line(25,0)(35,0)
\Line(0,25)(0,35)
\Line(0,-25)(0,-35)
\DashLine(-12,30)(-12,-30){3}
\Text(-30,0)[]{\footnotesize{$K$} }
\Text(10,+30)[]{\footnotesize{$K_r$} }
\Text(10,-30)[]{\footnotesize{$K_s$} }
\end{picture}
\end{center}
\vspace*{0.5cm}
\caption{Double-cut of Box functions}
\label{fig:BoxCoeff}
\end{figure}
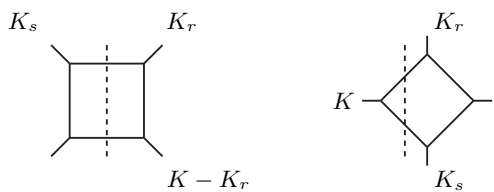

\noindent
The formula for the
coefficient of either of the box-functions with external
kinematics as shown in Fig.\ref{fig:BoxCoeff} is

\bea C[Q_r,Q_s,K] & = & {(K^2)^{2+n}\over 2}\left({\prod_{j=1}^{k+n}
\gb{P_{sr,1}|R_j |P_{sr,2}}\over \gb{P_{sr,1}|K
|P_{sr,2}}^{n+2}\prod_{t=1,t\neq i,j}^k \gb{P_{sr,1}|Q_t
|P_{sr,2}}}+ \{P_{sr,1}\leftrightarrow P_{sr,2}\}
\right),~~\Label{box-exp}\eea
where
\bea
\Delta_{sr} &=& (2Q_s \cdot Q_r)^2-4 Q_s^2 Q_r^2  \nonumber \\
P_{sr,1} &=& Q_s + \left( {-2Q_s \cdot Q_r + \sqrt{\Delta_{sr}}\over 2Q_r^2} \right) Q_r \nonumber \\
P_{sr,2} &=& Q_s + \left( {-2Q_s \cdot Q_r - \sqrt{\Delta_{sr}}\over
2Q_r^2} \right) Q_r~~~~\Label{box-null} \eea

\subsection{Triangle coefficient}

\vspace*{1cm}

\begin{figure}[h]
\begin{center}
\begin{picture}(0,0)(0,0)
\SetScale{0.7}
\SetWidth{1.0}
\Line(+20,-20)(-20,0)
\Line(+20,20)(-20,0)
\Line(+20,-20)(+20,20)
\Line(+20,-20)(+30,-30)
\Line(-20,0)(-30,10)
\Line(-20,0)(-30,-10)
\Line(+20,20)(+30,30)
\DashLine(0,30)(0,-30){3}
\Text(-30,0)[]{\footnotesize{$K$}}
\Text(+30,+30)[]{\footnotesize{$K_s$}}
\end{picture}
\end{center}
\vspace*{0.5cm}
\caption{Double-cut of a Triangle}
\label{fig:TriCoeff}
\end{figure}
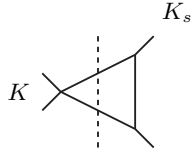

%
%
%
\noindent
The formula for the
coefficient of the triangle function with external
kinematics as shown in Fig.\ref{fig:TriCoeff} is
\bea C[Q_s,K] & = & { (K^2)^{1+n}\over
2}\frac{1}{(\sqrt{\Delta_s})^{n+1}}\frac{1}{(n+1)!
\vev{P_{s,1}~P_{s,2}}^{n+1}} \nonumber
\\ & & \times \frac{d^{n+1}}{d\tau^{n+1}}\left.\left({\prod_{j=1}^{k+n}
\vev{P_{s,1}-\tau P_{s,2} |R_j Q_s|P_{s,1}-\tau P_{s,2}}\over
\prod_{t=1,t\neq s}^k \vev{P_{s,1}-\tau P_{s,2}|Q_t Q_s
|P_{s,1}-\tau P_{s,2}}} + \{P_{s,1}\leftrightarrow
P_{s,2}\}\right)\right|_{\tau=0},~~~~~\Label{tri-exp}\eea
where
\bea
\Delta_{s} &=& (2Q_s \cdot K)^2-4 Q_s^2 K^2 \nonumber \\
P_{s,1} &=& Q_s + \left({-2Q_s \cdot K + \sqrt{\Delta_{s}}\over
2K^2} \right) K
\nonumber \\
P_{s,2} &=& Q_s + \left({-2Q_s \cdot K - \sqrt{\Delta_{s}}\over
2K^2} \right) K ~~~~\Label{tri-null} \eea
Note that the triangle coefficient is present only when $n\geq
-1$.

\subsection{Bubble coefficient}

\vspace*{1cm}

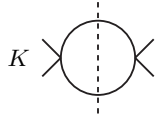
\begin{figure}[h]
\begin{center}
\begin{picture}(0,0)(0,0)
\SetScale{0.7}
\SetWidth{1.0}
\Line(-20,0)(-30,10)
\Line(-20,0)(-30,-10)
\Line(20,0)(30,10)
\Line(20,0)(30,-10)
\Oval(0,0)(20,20)(0)
\DashLine(0,30)(0,-30){3}
\Text(-30,0)[]{\footnotesize{$K$}}
\end{picture}
\end{center}
\vspace*{0.5cm}
\caption{Double-cut of a Bubble}
\label{fig:BubCoeff}
\end{figure}

\noindent
The formula for the
coefficient of the bubble function with the external
momentum $K$, shown in Fig.\ref{fig:BubCoeff}, is
\bea
 C[K] = (K^2)^{1+n} \sum_{q=0}^n {(-1)^q\over q!} {d^q \over
ds^q}\left.\left( {\cal B}_{n,n-q}^{(0)}(s)+\sum_{r=1}^k\sum_{a=q}^n
\left({\cal B}_{n,n-a}^{(r;a-q;1)}(s)-{\cal
B}_{n,n-a}^{(r;a-q;2)}(s)\right)\right)\right|_{s=0},~~~~~\Label{bub-exp}
\eea
where
\bea {\cal B}_{n,t}^{(0)}(s)\equiv {d^n\over d\tau^n}\left.\left( {1
\over n! [\eta|\W \eta K|\eta]^{n}}  {(2\eta\cdot K)^{t+1} \over
(t+1) (K^2)^{t+1}}{\prod_{j=1}^{n+k} \vev{\ell|R_j
(K+s\eta)|\ell}\over \vev{\ell~\eta}^{n+1} \prod_{p=1}^k \vev{\ell|
Q_p(K+s\eta)|\ell}}|_{\ket{\ell}\to |K-\tau \W \eta|\eta]
}\right)\right|_{\tau= 0},~~~\Label{cal-B-0}\eea
\bea & & {\cal B}_{n,t}^{(r;b;1)}(s)  \equiv  {(-1)^{b+1}\over
 b! \sqrt{\Delta_r}^{b+1} \vev{P_{r,1}~P_{r,2}}^b}{d^b \over d\tau^{b}}
\left({1\over (t+1)} {\gb{P_{r,1}-\tau
P_{r,2}|\eta|P_{r,1}}^{t+1}\over \gb{P_{r,1}-\tau
P_{r,2}|K|P_{r,1}}^{t+1}}\right. \nonumber \\ & & \times
\left.\left. {\vev{P_{r,1}-\tau P_{r,2}|Q_r \eta|P_{r,1}-\tau
P_{r,2}}^{b} \prod_{j=1}^{n+k} \vev{P_{r,1}-\tau P_{r,2}|R_j
(K+s\eta)|P_{r,1}-\tau P_{r,2}}\over \vev{P_{r,1}-\tau P_{r,2}|\eta
K|P_{r,1}-\tau P_{r,2}}^{n+1} \prod_{p=1,p\neq r}^k
\vev{P_{r,1}-\tau P_{r,2}| Q_p(K+s\eta)|P_{r,1}-\tau
P_{r,2}}}\right)\right|_{\tau=0},~~~\Label{cal-B-r-1}\eea
\bea & & {\cal B}_{n,t}^{(r;b;2)}(s)  \equiv  {(-1)^{b+1}\over
 b! \sqrt{\Delta_r}^{b+1} \vev{P_{r,1}~P_{r,2}}^{b}}{d^{b} \over d\tau^{b}}
\left({1\over (t+1)} {\gb{P_{r,2}-\tau
P_{r,1}|\eta|P_{r,2}}^{t+1}\over \gb{P_{r,2}-\tau
P_{r,1}|K|P_{r,2}}^{t+1}}\right. \nonumber \\ & & \times
\left.\left. {\vev{P_{r,2}-\tau P_{r,1}|Q_r \eta|P_{r,2}-\tau
P_{r,1}}^{b} \prod_{j=1}^{n+k} \vev{P_{r,2}-\tau P_{r,1}|R_j
(K+s\eta)|P_{r,2}-\tau P_{r,1}}\over \vev{P_{r,2}-\tau P_{r,1}|\eta
K|P_{r,2}-\tau P_{r,1}}^{n+1} \prod_{p=1,p\neq r}^k
\vev{P_{r,2}-\tau P_{r,1}| Q_p(K+s\eta)|P_{r,2}-\tau
P_{r,1}}}\right)\right|_{\tau=0}.~~~\Label{cal-B-r-2}\eea
where $\Delta_r, P_{r,1}, P_{r,2}$ are given by (\ref{tri-null}), and $\eta, \W\eta$ are arbitrary, generically chosen null vectors.
Note that the bubble coefficient exists only when $n\geq 0$.

%
%

\section{Example I: $s_{12}$-channel cut of $A(1^+,2^+,3^+,H)$}

In this section as well as the next, we check our formulas by
reconstructing some helicity amplitudes contributing to $gH \to gg$
and $gg \to gg$ at NLO in QCD, both known in the literature
\cite{Ellis:1987xu,Bern:1995db}.  We present our calculations in detail.

Our first example is the
$s_{12}$-channel cut of $A(1^+,2^+,3^+,H)$.  This amplitude
was first computed in \cite{Ellis:1987xu}.  Here, to facilitate comparison, we follow the
setup of \cite{Rozowsky:1997dm}, where the amplitude was rederived
using unitarity cuts.  At one loop, every Feynman diagram has a massive quark circulating in the loop.  The quark mass is denoted by $m$.

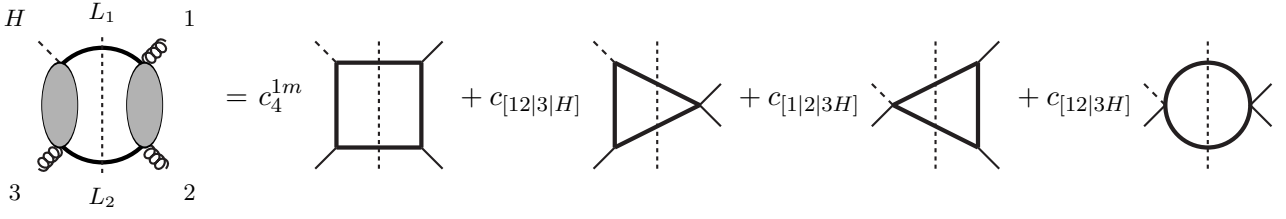
\begin{figure}
\bea
\begin{picture}(0,0)(0,0)
\SetScale{0.8}
\SetWidth{2.0}
\GOval(0,0)(27,27)(0){1}
\SetWidth{1.0}
\Gluon(-20,-20)(-30,-30){3}{3}
\Gluon(20,-20)(30,-30){3}{3}
\Gluon(20,20)(30,30){3}{3}
\DashLine(-20,20)(-30,30){3}
\GOval(-20,0)(20,8.5)(0){0.7}
\GOval(+20,0)(20,8.5)(0){0.7}
\DashLine(0,32)(0,-32){2}
\Text(-33, -33)[]{\footnotesize {$3$}}
\Text(-33, +33)[]{\footnotesize {$H$}}
\Text(+33, +33)[]{\footnotesize {$1$}}
\Text(+33, -33)[]{\footnotesize {$2$}}
\Text(0, +35)[]{\footnotesize {$L_1$}}
\Text(0, -35)[]{\footnotesize {$L_2$}}
\end{picture}
\hspace*{1.5cm} &=&
c_4^{1m} \hspace*{1cm}
\begin{picture}(0,0)(0,0)
\SetScale{0.8}
\SetWidth{2.0}
\Line(-20,-20)(20,-20)
\Line(20,-20)(20,20)
\Line(20,20)(-20,20)
\Line(-20,20)(-20,-20)
\SetWidth{1.0}
\Line(-20,-20)(-30,-30)
\Line(20,-20)(30,-30)
\Line(20,20)(30,30)
\DashLine(-20,20)(-30,30){3}
\DashLine(0,30)(0,-30){2}
\end{picture}
\hspace*{1cm} +
c_{[12|3|H]}\hspace*{1cm}
\begin{picture}(0,0)(0,0)
\SetScale{0.8}
\SetWidth{2.0}
\Line(-20,-20)(20,0)
\Line(-20,20)(20,0)
\Line(-20,-20)(-20,20)
\SetWidth{1.0}
\Line(-20,-20)(-30,-30)
\Line(20,0)(30,10)
\Line(20,0)(30,-10)
\DashLine(-20,20)(-30,30){3}
\DashLine(0,30)(0,-30){2}
\end{picture}
\hspace*{1cm} +
c_{[1|2|3H]}\hspace*{1cm}
\begin{picture}(0,0)(0,0)
\SetScale{0.8}
\SetWidth{2.0}
\Line(+20,-20)(-20,0)
\Line(+20,20)(-20,0)
\Line(+20,-20)(+20,20)
\SetWidth{1.0}
\Line(+20,-20)(+30,-30)
\DashLine(-20,0)(-30,10){3}
\Line(-20,0)(-30,-10)
\Line(+20,20)(+30,30)
\DashLine(0,30)(0,-30){2}
\end{picture}
\hspace*{1cm} +
c_{[12|3H]}\hspace*{1cm}
\begin{picture}(0,0)(0,0)
\SetScale{0.8}
\SetWidth{2.0}
\Oval(0,0)(20,20)(0)
\SetWidth{1.0}
\DashLine(-20,0)(-30,10){3}
\Line(-20,0)(-30,-10)
\Line(20,0)(30,10)
\Line(20,0)(30,-10)
\DashLine(0,30)(0,-30){2}
\end{picture}
\nonumber
\eea
\vspace*{0.5cm}
\caption{Double-cut in the $s_{12}$-channel for $A(1^+,2^+,3^+,H)$.}
\label{fig:s12cut}
\end{figure}

The $s$-channel cut of $A(1^+,2^+,3^+,H)$ admits a decomposition
in terms of cuts of master integrals as shown in Fig. \ref{fig:s12cut}.
Its expression, given in Eq.(4.20) of \cite{Rozowsky:1997dm}, reads
\bea
A(1^+,2^+,3^+,H)|_{s-{\rm channel}}&=&
- { i \over (4 \pi)^{2 - \epsilon} }
{m^2 \over v^2}
{ s t \over \spa 1.2 \spa 2.3 \spa 3.1}
\Bigg\{
  {(t-u) \over 2 t (s-m_H^2)} I_3^{(4)}[4 (m^2 +\mu^2) - s] \nn
&& \qquad
- {(s-m_H^2) \over 2 s t} I_3^{(2)}[4 (m^2 +\mu^2) - s]
- {1 \over 2} I_4[4 (m^2 +\mu^2) - s] \ .
\Bigg\}
\eea
where $s=s_{12}, \ t=s_{23} \ , u=m_H^2 - s - t$.
One can thus read the following values for the coefficients:
\bea
c_4^{1m} &=& - c_0 \ {1 \over 2} \Big(4 (m^2 +\mu^2) - s \Big) \ , \\
c_{[12|3|H]} &=& - c_0 \ {(s-m_H^2) \over 2 s t} \Big(4 (m^2 +\mu^2) - s \Big) \ , \\
c_{[1|2|3H]} &=& c_0 \ {(t-u) \over 2 t (s-m_H^2)} \Big(4 (m^2 +\mu^2) - s \Big) \ , \\
c_{[12|3H]} &=& 0 ~~~\Label{exa-1-bubble}\ . \eea with \bea c_0 &=&
- { i \over (4 \pi)^{2 - \epsilon} } {m^2 \over v^2} { s t \over
\spa 1.2 \spa 2.3 \spa 3.1} \ . \eea

\subsection{The reconstruction of the coefficients}

We now show how to reconstruct the coefficients given above with our formulas from Section \ref{sec:formulas}.
We follow the definition of the integrand given by \cite{Rozowsky:1997dm}.
By sewing the tree-level amplitude $A_4^{\rm tree}(-L_1,1^+,2^+,-L_2)$ and
$A_4^{\rm tree}(-L_1,3^+,H,-L_2)$ given in Eq.(4.1-4.2) of \cite{Rozowsky:1997dm},
and using the Dirac equation for massive fermion,
it is shown that
the four-dimensional integrand of the $s$-cut, $C_{12}$, can be written as:\footnote{Note that here we use ``twistor'' sign convention for the antiholomorphic spinor product, which is the opposite of the ``QCD'' convention followed by \cite{Rozowsky:1997dm}
$\spb {x}.{y}^{\rm Rozowsky} = - \spb {x}.{y}^{\rm BFM} $.
}
\bea
C_{12} = c_{0,1} {N_1 \over D_2} + c_{0,2}{N_2 \over D_2 D_4}
\eea
where
\bea
c_{0,1}=-{ m K^2 s_{23} \over v (K^2-m_H^2) \vev{1~2}\vev{2~3}\vev{3~1}}
 ~~~~~c_{0,2}=
{m \cb{1~2} \over \sqrt{2} v \vev{1~2}}
\eea
\bea
N_1 &=& m \left[ 4(m^2 + \mu^2)-s_{12}\right] \\
N_2 &=& 8m(m^2 +\mu^2)(\ell_1 \cdot \eps_3^+ + k_4 \cdot \eps_3^+)
+\sqrt{2} m {\gb{1|4|\ell_1|4|3} \over \vev{1~3}}
\eea
\bea D_2 & = & (\ell_1-k_1)^2 - \mu^2 - m^2, ~~~D_4 = (\ell_1+k_4)^2
- \mu^2 - m^2 \eea
\noindent
We need to classify the contribution of $N_1$ and $N_2$ to each
coefficient.  Observe that  $N_1$ is independent of the loop momentum variable, so we consider it as a single term,
\bea N_1
&=& m \left[ 4(m^2 + \mu^2)-s_{12}\right], \eea
while $N_2$ is treated as three separate terms,
\bea N_2 = N_{2,1} + N_{2,2} +
N_{2,3} \, \eea where 
\bea N_{2,1} &=& 8m(m^2 +\mu^2)(k_4 \cdot
\eps_3^+)
         = 8m(m^2 +\mu^2) {\spab{1}.{2}.{3} \over \sqrt{2} \spa 1.3 } \\
N_{2,2} &=& 8m(m^2 +\mu^2)(\ell_1 \cdot \eps_3^+)
         = - 8m(m^2 +\mu^2) {\spab{1}.{\ell_1}.{3} \over \sqrt{2} \spa 1.3 } \\
N_{2,3} &=& \sqrt{2} m {\gb{1|4|\ell_1|4|3} \over \vev{1~3}}.
\eea
By pattern-matching onto the
reference form in Eq.(\ref{frame}),
each integrand can be characterised by the
parameters given in the following table.

\bea
\begin{array}{|c||c|c|c|}
\hline
{\rm integrand} & n & k & \s{P}_1 = |P_1\rangle [P_1| \\
\hline
\hline
{N_1 / D_2} & -1 & 1 & - \\
\hline
{N_{2,1} /(D_2 D_4)} & -2 & 2 &  - \\
\hline
{N_{2,2} /(D_2 D_4)} & -1 & 2 & |1 \rangle \ [3|\\
\hline
{N_{2,3} /(D_2 D_4)} & -1 & 2 &  k_4|3] \ \langle 1|k_4 \\
\hline
\end{array}
\eea
These data are  the input values that we need in evaluating the formulas of the coefficients of the master integrals. \\

From this table we draw the following conclusions. Since $N_1/D_2$
has $k=1$ and $n=-1$, it contributes only to a triangle coefficient;
whereas $N_{2,i} \ (i=1,2,3)$, having $n\leq -1$, contributes to both box
and triangle coefficients. There are no bubble contributions at all. Thus we
have already reproduced the absence of bubbles, Eq.(\ref{exa-1-bubble}), without any
calculation.


To apply our formulas, we need to identify the definitions of $K$, $K_1$, and $K_2$.  By inspection of $D_2$ and $D_4$, along with the fact of working in the $s$-channel cut, we choose the following consistent definitions:

\bea
K=k_1+k_2;~~~K_1=k_1;~~~K_2=-k_4.
~~~\Label{ks-for-gggh}
\eea
Since there is a single massive quark circulating in the loop, we have
\bea
~~~ M_1=M_2=m_j=m,\eea
and
\bea z= {1-\sqrt{1-u}\sqrt{1-{4 m^2\over s_{12}}}\over 2} \eea
From (\ref{ks-for-gggh}), we use the definition (\ref{Q-def}) to construct
 \bea
Q_1 &=& -(1-z)k_1 - z k_2 \\
Q_2 
&=& \left(1-2z \right)k_4 +\left((1-z){m_H^2
    \over  K^2} - z  \right)K
\eea
Using (\ref{tri-null}), we also set up the following quantities useful for triangle coefficients:
\bea & & \Delta_1 = (1-2z)^2 (K^2)^2 \\
& & P_{1,1} = (1-2z)k_2,~~~~~P_{1,2}=-(1-2z)k_1\\& & \Delta_2 = (1-2z)^2 (K^2-m_H^2)^2 \\
& & P_{2,1} = -(1-2z)k_3,~~~~~ P_{2,2}=-(1-2z){m_H^2 \over
K^2}k_3+(1-2z)(1-{m_H^2 \over K^2})k_4\eea
%



\subsection{The box coefficient  $c_4^{1m}$}

The box coefficient $c_4^{1m}$ takes contributions from
$N_{2,1}$, $N_{2,2}$, $N_{2,3}$:
\bea
c_4^{1m} 
&=& c_{0,2}~8m(m^2+\mu^2)k_4 \cdot \eps_3^+
\ C[Q_1,Q_2,K]^{(2,1)} - c_{0,2} {8m(m^2+\mu^2) \over
\sqrt{2}\vev{1~3}} \ C[Q_1,Q_2,K]^{(2,2)} \nn && + c_{0,2} {\sqrt{2}
m \over \vev{1~3}} \ C[Q_1,Q_2,K]^{(2,3)} \eea
where $C[Q_r,Q_s,K]$, defined in Eq.(\ref{box-exp}), is
\bea C[Q_r,Q_s,K] & = & {(K^2)^{2+n}\over 2}\left({\prod_{j=1}^{k+n}
\gb{P_{sr,1}|R_j |P_{sr,2}}\over \gb{P_{sr,1}|K
|P_{sr,2}}^{n+2}\prod_{t=1,t\neq r,s}^k \gb{P_{sr,1}|Q_t
|P_{sr,2}}}+ \{P_{sr,1}\leftrightarrow P_{sr,2}\}
\right).
\eea

\bigskip

\noindent $\bullet \ C[Q_1,Q_2,K]^{(2,1)}$

This term, corresponding to $n=-2$ is trivial, since $k=2$ and $N_{2,1}$ has no dependence on the
loop variable,
\bea
C[Q_1,Q_2,K]^{(2,1)} = 1
\eea

\bigskip

\noindent $\bullet \ C[Q_1,Q_2,K]^{(2,2)}$ and $C[Q_1,Q_2,K]^{(2,3)}$

$C[Q_1,Q_2,K]^{(2,2)}$ and $C[Q_1,Q_2,K]^{(2,3)}$ both correspond to
$n=-1,k=2$. They differ only in the definition of $P_1$. Therefore we
can compute them in parallel, and specialize later to the corresponding
$P_1$. With $n=-1,k=2$, the
expression is
\bea & & C[Q_1,Q_2,K]  =  {(K^2)\over 2}\left({
\gb{P_{21,1}|R_1 |P_{21,2}}\over \gb{P_{21,1}|K
|P_{21,2}} }+ \{P_{21,1}\leftrightarrow P_{21,2}\}
\right) \\
 & & =  -(1-2z){K^2\over 2}
\left({\gb{P_{21,1}|P_1 |P_{21,2}}\over \gb{P_{21,1}|K
|P_{21,2}} }
+ {\gb{P_{21,2}|P_1 |P_{21,1}}\over \gb{P_{21,2}|K
|P_{21,1}} }
\right) + {z (-2K \cdot P_1) } \nonumber  \\
&  & =  -(1-2z){K^2\over 2}
\left({\gb{P_{21,1}|P_1 |P_{21,2}}\gb{P_{21,2}|K
|P_{21,1}}+\gb{P_{21,2}|P_1 |P_{21,1}}\gb{P_{21,1}|K
|P_{21,2}}\over \gb{P_{21,1}|K
|P_{21,2}}\gb{P_{21,2}|K
|P_{21,1}} }
\right) + {z (-2K \cdot P_1) } \nonumber
\eea

\bigskip

\noindent
For
$|P_1\rangle = |1\rangle \ , |P_1] = |3]$,
(so that, for any $S$, one has $2P_1 \cdot S = -\gb{1|S|3}$),
one obtains
\bea & & C[Q_1,Q_2,K]^{(2,2)}  =  \nonumber \\
& & = -(1-2z){K^2\over 2}
\left(-{(1-2z)^4 \over z(1-z)}{s_{23} (s_{23}+K^2-m_H^2)\gb{1|2|3} \over K^2}
\right)
\left({(1-2z)^4 \over z(1-z)}s_{23}(s_{23}+K^2-m_H^2)
\right)^{-1}
+ {z \gb{1|2|3} } \nonumber \\
& & ={\gb{1|2|3}\over 2}
\eea
\\
For
$|P_1\rangle = \s{k}_4|3] \ , |P_1] = \s{k}_4|1\rangle$,
( so that, $2P_1 \cdot S = -\gb{1|k_4 S k_4|3}$),
one gets
\bea & & C[Q_1,Q_2,K]^{(2,3)}  =  \nonumber \\
& & = -(1-2z){K^2\over 2}
\left({\gb{P_{21,1}|P_1 |P_{21,2}}\gb{P_{21,2}|K
|P_{21,1}}+\gb{P_{21,2}|P_1 |P_{21,1}}\gb{P_{21,1}|K
|P_{21,2}}\over \gb{P_{21,1}|K
|P_{21,2}}\gb{P_{21,2}|K
|P_{21,1}} }
\right) + {z (-2K \cdot P_1) } \nonumber
 \\
& & = -{\gb{1|2|3}\over 2}m_H^2
\eea

\bigskip

\noindent $\bullet \ $ The result for $c_4^{1m}$

The total coefficient of our box is:
\bea
c_4^{1m} &=&
  c_{0,2}~8m(m^2+\mu^2)k_4 \cdot \eps_3^+ -
c_{0,2} {8m(m^2+\mu^2) \over \sqrt{2}\vev{1~3}}({\gb{1|2|3}\over 2}) +
c_{0,2} {\sqrt{2} m \over \vev{1~3}}(-m_H^2 {\gb{1|2|3}\over 2}) \nonumber  \\
& = & - {m^2 K^2 s_{23} \over 2 v \vev{1~2}\vev{2~3}\vev{3~1}}
\left[
 {4(m^2+\mu^2) }
-m_H^2  \right]
\eea
Multiplying by $-i/(4\pi)^{2-\eps}$, to account for the difference in the definitions of master integrals, we confirm the result of \cite{Rozowsky:1997dm}.

\subsection{The triangle coefficient  $c_{[1|2|3H]}$}

The coefficient $c_{[1|2|3H]}$ gets contributions from
$N_{1}$, $N_{2,2}$ and $N_{2,3}$:
\bea
c_{[1|2|3H]} & = & c_{0,1} \ N_1 \ C[Q_1,K]^{(1)}
- c_{0,2} {8m(m^2+\mu^2) \over \sqrt{2}\vev{1~3}} \ C[Q_1,K]^{(2,2)}
+ c_{0,2} {\sqrt{2} m \over \vev{1~3}} \ C[Q_1,K]^{(2,3)}
\eea
where the general triangle coefficient, given in Eq.(\ref{tri-exp}),
reads
\bea
C[Q_s,K] & = & { (K^2)^{1+n}\over
2}\frac{1}{(\sqrt{\Delta_s})^{n+1}}\frac{1}{(n+1)!
\vev{P_{s,1}~P_{s,2}}^{n+1}} \nonumber
\\ & & \times \frac{d^{n+1}}{d\tau^{n+1}}\left.\left({\prod_{j=1}^{k+n}
\vev{P_{s,1}-\tau P_{s,2} |R_j Q_s|P_{s,1}-\tau P_{s,2}}\over
\prod_{t=1,t\neq s}^k \vev{P_{s,1}-\tau P_{s,2}|Q_t Q_s
|P_{s,1}-\tau P_{s,2}}} + \{P_{s,1}\leftrightarrow
P_{s,2}\}\right)\right|_{\tau=0}.
\eea

\bigskip

\noindent $\bullet \ C[Q_1,K]^{(1)}$

We have already observed that the $N_1$ term is trivial.
Here is how that shows up in our formulas.

Read $C[Q_s,K]$ for $s=1$ and $k=1,n=-1.$, and no $R_j$.

The term inside the parentheses degenerates to 1.
\bea
C[Q_1,K]^{(1)} & = & { 1 \over
2}  \left.\left(1 +1 \right)\right|_{\tau=0} = 1 .
\eea

\bigskip

\noindent $\bullet \ C[Q_1,K]^{(2,2)}$ and $C[Q_1,K]^{(2,3)}$

As we said already, $N_{2,2}$ and $N_{2,3}$ differ in the
definition of $P_1$. Therefore we start by manipulating the general formula, and only at the very end
we specialize each contribution using the corresponding $P_1$.

Since $n=-1,k=2$, there is no derivative at all, so we can set
$\tau=0$ from the beginning:
\bea
C[Q_s,K] & = & { 1\over
2}  \left({
\vev{P_{s,1} |R_1 Q_s|P_{s,1}}\over
\prod_{t=1,t\neq s}^k \vev{P_{s,1}|Q_t Q_s
|P_{s,1}}} +
{\vev{P_{s,2} |R_1 Q_s|P_{s,2}}\over
\prod_{t=1,t\neq s}^k \vev{P_{s,2}|Q_t Q_s
|P_{s,2}}}
\right).
\eea

The one-mass triangle $(1|2|3H)$ corresponds to the value $s=1$.
\bea
C[Q_1,K]
 & = & -{ 1\over2}  \left({
\gb{2 |P_1|1}\over \gb{2|4|1}} +
{\gb{1 |P_1|2}\over \gb{1|4|2}}
\right)
\eea
For
$|P_1\rangle = |1\rangle \ , |P_1] = |3]$,
one gets
\bea
C[Q_1,K]^{(2,2)} & = & -{ 1\over2}  {
\gb{1|2|3} \over s_{23} }
\eea
For
$|P_1\rangle = \s{k}_4|3] \ , |P_1] = \s{k}_4|1\rangle$,
one obtains
\bea C[Q_1,K]^{(2,3)} & = & -{ 1\over2}  \left({
\gb{2|4|3}\gb{1|4|1}\over \gb{2|4|1}} + {\gb{1|4|3}\gb{1|4|2}\over
\gb{1|4|2}} \right) =\gb{1|2|3}\left(1-{m_H^2 \over 2 s_{23}}
\right) \eea

\bigskip

\noindent $\bullet \ $ The result for $c_{[1|2|3H]}$

The total coefficient of triangle $(1|2|3H)$ is:
\bea
c_{[1|2|3H]} &= & -{ m K^2 s_{23} \over (K^2-m_H^2) \vev{1~2}\vev{2~3}\vev{3~1}}  m \left[ 4(m^2 + \mu^2)-s_{12}\right] -
{m \cb{1~2} \over \sqrt{2} v \vev{1~2}} {8m(m^2+\mu^2) \over
  \sqrt{2}\vev{1~3}}(-{\gb{1|2|3}\over 2 s_{23}}) \nonumber \\ & & +
{m \cb{1~2} \over \sqrt{2} v \vev{1~2}} {\sqrt{2} m \over \vev{1~3}}\gb{1|2|3}\left(1-{m_H^2
\over 2 s_{23}} \right)\nonumber  \\
&= &
-{m^2 K^2 \left(2 s_{23}+K^2-m_H^2\right)\over 2 v (K^2-m_H^2)\vev{1~2} \vev{2~3}\vev{3~1}}
  \left[ 4(m^2 + \mu^2)-m_H^2\right]
\eea
Multiplying by $i/(4\pi)^{2-\eps}$, to account for the difference in the definitions of master integrals, we again confirm the result of \cite{Rozowsky:1997dm}.

\subsection{The coefficient $c_{[12|3|H]}$}

The coefficient $c_{[12|3|H]}$ gets contributions from
$N_{2,2}$ and $N_{2,3}$, therefore
can be written as
\bea
c_{[12|3|H]} & = &
- c_{0,2} {8m(m^2+\mu^2) \over \sqrt{2}\vev{1~3}} \ C[Q_2,K]^{(2,2)}
+ c_{0,2} {\sqrt{2} m \over \vev{1~3}} \ C[Q_2,K]^{(2,3)}
\eea

\bigskip

\noindent $\bullet \ C[Q_2,K]^{(2,2)}$ and $C[Q_2,K]^{(2,3)}$

The two-mass triangle $(12|3|H)$ corresponds to the value $s=2$,
and its coefficient can be obtained from
 Eq.(\ref{tri-exp}),
\bea
C[Q_2,K] 
 & = & { 1\over
2}  \left({
\vev{3 |P_1 K |3}
\over\vev{3|1|2|3}
}-
{  \cb{3|K P_1|3}
\over
  \cb{3|1|2|3} }
\right)
\eea
By using
$|P_1\rangle = |1\rangle \ , |P_1] = |3]$,
one gets
\bea
C[Q_2,K]^{(2,2)}
 & = &
\left( 1-{m_H^2 \over K^2} \right)
{\gb{1|2|3} \over 2  s_{23}}
\eea
By using
$|P_1\rangle = \s{k}_4|3] \ , |P_1] = \s{k}_4|1\rangle$,
one obtains
\bea
C[Q_2,K]^{(2,3)}
 & = &
 { 1\over
2}  \left({
\gb{3|4|3}\vev{1|4  K |3}
\over\vev{3|1|2|3}
}-
{  \cb{3|K4|3}\gb{1|4|3}
\over
  \cb{3|1|2|3} }
\right) = (1-{m_H^2 \over K^2}) m_H^2 {\gb{1|2|3} \over 2 s_{23}}
\eea

\bigskip

\noindent $\bullet \ $ The result of $c_{[12|3|H]}$

The total coefficient of triangle $(12|3|H)$ is:
\bea
c_{[12|3|H]} & = &
-c_{0,2} {8m(m^2+\mu^2) \over \sqrt{2}\vev{1~3}}\left( 1-{m_H^2 \over K^2}\right) {\gb{1|2|3} \over 2  s_{23}}
 +
c_{0,2} {\sqrt{2} m \over \vev{1~3}}\left( 1-{m_H^2 \over K^2}\right)
m_H^2 {\gb{1|2|3} \over 2 s_{23}}
 \nonumber \\
& & =
{m^2 (K^2-m_H^2)  \over 2  v \vev{1~2} \vev{2~3}\vev{3~1}}
\left[
{4(m^2+\mu^2) }-m_H^2
\right]
\eea
Multiplying by $i/(4\pi)^{2-\eps}$, to account for the difference in the definitions of master integrals, we again confirm the result of \cite{Rozowsky:1997dm}.

\section{Example II: $s_{23}$-channel cut of $A(1^-,2^-,3^+,4^+)$}

Our second example features a non-vanishing bubble coefficient.  We study the $t$-channel cut of the gluon amplitude  $A(1^-,2^-,3^+,4^+)$, with a massive quark circulating in the loop.  (As usual, $t=s_{23}$.)

\begin{figure}
\bea
\begin{picture}(0,0)(0,0)
\SetScale{0.8}
\SetWidth{2.0}
\GOval(0,0)(27,27)(0){1}
\SetWidth{1.0}
\Gluon(-20,-20)(-30,-30){3}{3}
\Gluon(20,-20)(30,-30){3}{3}
\Gluon(20,20)(30,30){3}{3}
\Gluon(-20,20)(-30,30){3}{3}
\GOval(-20,0)(20,8.5)(0){0.7}
\GOval(+20,0)(20,8.5)(0){0.7}
\DashLine(0,32)(0,-32){2}
\Text(-33, -33)[]{\footnotesize {$4$}}
\Text(-33, +33)[]{\footnotesize {$1$}}
\Text(+33, +33)[]{\footnotesize {$2$}}
\Text(+33, -33)[]{\footnotesize {$3$}}
\Text(0, +35)[]{\footnotesize {$L_1$}}
\Text(0, -35)[]{\footnotesize {$L_2$}}
\end{picture}
\hspace*{1.5cm} &=&
c_4^{0m} \hspace*{1cm}
\begin{picture}(0,0)(0,0)
\SetScale{0.8}
\SetWidth{2.0}
\Line(-20,-20)(20,-20)
\Line(20,-20)(20,20)
\Line(20,20)(-20,20)
\Line(-20,20)(-20,-20)
\SetWidth{1.0}
\Line(-20,-20)(-30,-30)
\Line(20,-20)(30,-30)
\Line(20,20)(30,30)
\Line(-20,20)(-30,30)
\DashLine(0,30)(0,-30){2}
\end{picture}
\hspace*{1cm} +
c_{[23|4|1]}\hspace*{1cm}
\begin{picture}(0,0)(0,0)
\SetScale{0.8}
\SetWidth{2.0}
\Line(-20,-20)(20,0)
\Line(-20,20)(20,0)
\Line(-20,-20)(-20,20)
\SetWidth{1.0}
\Line(-20,-20)(-30,-30)
\Line(20,0)(30,10)
\Line(20,0)(30,-10)
\Line(-20,20)(-30,30)
\DashLine(0,30)(0,-30){2}
\end{picture}
\hspace*{1cm} +
c_{[2|3|41]}\hspace*{1cm}
\begin{picture}(0,0)(0,0)
\SetScale{0.8}
\SetWidth{2.0}
\Line(+20,-20)(-20,0)
\Line(+20,20)(-20,0)
\Line(+20,-20)(+20,20)
\SetWidth{1.0}
\Line(+20,-20)(+30,-30)
\Line(-20,0)(-30,10)
\Line(-20,0)(-30,-10)
\Line(+20,20)(+30,30)
\DashLine(0,30)(0,-30){2}
\end{picture}
\hspace*{1cm} +
c_{[12|34]}\hspace*{1cm}
\begin{picture}(0,0)(0,0)
\SetScale{0.8}
\SetWidth{2.0}
\Oval(0,0)(20,20)(0)
\SetWidth{1.0}
\Line(-20,0)(-30,10)
\Line(-20,0)(-30,-10)
\Line(20,0)(30,10)
\Line(20,0)(30,-10)
\DashLine(0,30)(0,-30){2}
\end{picture}
\nonumber
\eea
\vspace*{0.5cm}
\caption{Double-cut in the $s_{23}$-channel for $A(1^-,2^-,3^+,4^+)$.}
\label{fig:s23cut}
\end{figure}
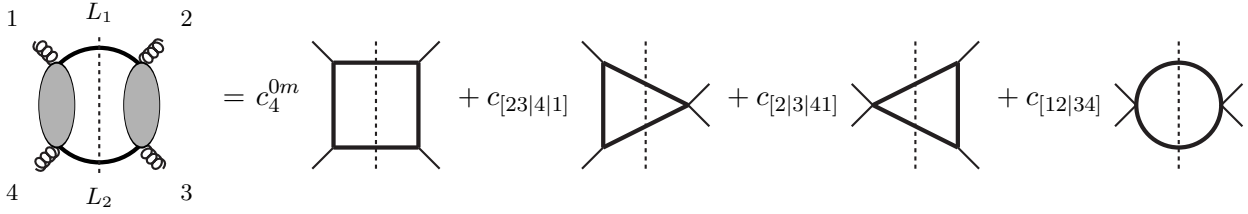

The $t$-channel cut of $A(1^-,2^-,3^+,4^+)$  admits a decomposition
in terms of cuts of master integrals as shown in Fig. \ref{fig:s23cut}, and
its expression was given in Eq.(5.33) of \cite{Bern:1995db}.
After converting that expression into our basis of $D$-dimensional master integrals, as done in
Appendix \ref{sec:IntegralsTranslation}, it reads
\bea
\left. A^{\rm fermion}_4(1^-,2^-,3^+,4^+) \right|_{t-{\rm cut}}
&=&
  \left. {\vev{1~2}^2 \cb{3~4}^2 \over st}
\left(
 {2 \over 3} I_2[1]
+{4 \over 3t}I_2[m^2+\mu^2]
- {2\over s} I_2[m^2+\mu^2]
\nonumber \right. \right. \\ & & \left. \left.
+{2t \over s} I_4[(m^2+\mu^2)^2]
- t  I_4[m^2+\mu^2]
\right) \right|_{t-{\rm cut}}
\eea
One reads the following values for the coefficients:
\bea
c_4^{0m} &=&
c_0 \ \bigg( {2t \over s} (m^2+\mu^2) - t \bigg) (m^2+\mu^2) \ , \\
c_{[23|4|1]} &=& 0 \ , \\
c_{[2|3|41]} &=& 0 \ , \\
c_{[23|41]} &=& c_0
\ \bigg(
 {2 \over 3}
+{4 \over 3t} (m^2+\mu^2)
- {2\over s} (m^2+\mu^2)
\bigg)
\eea
with
\bea
c_0 &=& {\vev{1~2}^2 \cb{3~4}^2 \over st} \ .
\eea

\subsection{The reconstruction of the coefficients}

We now apply our formulas of Section \ref{sec:formulas} to construct the coefficients given above.
We follow the definition of the integrand given by \cite{Bern:1995db}.
By sewing the tree-level amplitude $A_4^{\rm tree}(-L_1,2^-,3^+,L_2)$ and
$A_4^{\rm tree}(-L_2,4^+,1^-,L_1)$ given in Eq.(2.3) of \cite{Bern:1995db},
and using the Dirac equation for a massive fermion,
it is shown that
the four-dimensional integrand of the $t$-cut, $C_{23}$, can be written as:\footnote{Recall that here we use ``twistor'' sign convention for the antiholomorphic spinor product, which is the opposite of the ``QCD'' convention followed by \cite{Bern:1995db}
$\spb {x}.{y}^{\rm Bern-Morgan} = - \spb {x}.{y}^{\rm BFM} $.
}%
\bea
C_{23} = -{2 N_1 + N_2 \over D_1 D_2},
\eea
with
\bea
N_1 &=& {1 \over s_{23}^2} \gb{1|\ell_1|4}^2 \gb{2|\ell_1|3}^2 \\
N_2 &=& -{1 \over s_{23}}\vev{1~2}\cb{3~4}\gb{1|\ell_1|4}\gb{2|\ell_1|3} \\
D_1 &=& (\ell_1+k_1)^2 - \mu^2 - m^2 \\
D_2 &=& (\ell_1-k_2)^2 - \mu^2 - m^2
\eea

By pattern-matching onto the
reference form in Eq.(\ref{frame}),
each integrand is characterised by the
parameters given in the following table.
\bea
\begin{array}{|c||c|c|c|c|c|c|}
\hline
{\rm integrand} & n & k &
   |P_1\rangle [P_1| & |P_2\rangle [P_2| & |P_3\rangle [P_3| & |P_4\rangle [P_4| \\
\hline
\hline
\hline
{N_{1} /(D_1 D_2)} & 2 & 2 & |1\rangle [4| & |2\rangle [3|  & |1\rangle [4| & |2\rangle [3| \\
\hline
{N_{2} /(D_1 D_2)} & 0 & 2 & |1\rangle [4| & |2\rangle [3|  & - & - \\
\hline
\end{array}
\eea
%


We define
\bea
& & K = k_2+k_3;~~~~~~~K_1=-k_1;~~~~~~K_2=k_2 \\
& & P_1=P_3 = \lambda_1\tilde\lambda_4,~~~~~~~P_2=P_4 =
\lambda_2\tilde\lambda_3.
\eea
Moreover, since we have a quark of mass $m$ circulating in the loop, we take
\bea
 M_1=M_2=m_j=m.
\eea
Then, by applying (\ref{solve-z}), we find
\bea
z(1-z) = {m^2 + \mu^2 \over K^2}.
~~~~
\Label{z-mu-ex2}
\eea
For the $N_1$ term, $n=2$.  For the $N_2$ term, $n=0$. Both terms give boxes, triangles and bubbles.

From the definitions (\ref{R-def}), (\ref{Q-def}) we have
\bea Q_1 &=& (1-z)k_1+z k_4,~~~Q_2 = -(1-z)k_2-z k_3 \\
R_1 &=& R_3 = -(1-2z)\lambda_1\tilde\lambda_4,~~~~ R_2 = R_4 =
-(1-2z)\lambda_2\tilde\lambda_3.
\eea
Further, the quantities defined in (\ref{box-null}), (\ref{tri-null}) become
\bea
\Delta_{12}  
&=&   (1-2z)^4 s_{12}(K^2+s_{12})  -(1-2z)^2 K^2   s_{12} \\
 \Delta_1 & = & (1-2z)^2 (K^2)^2,~~~\\P_{1,1} &=& -(1-2z)k_4,~~~
P_{1,2} = (1-2z)k_1,\\
\Delta_2 &=&  (1-2z)^2 (K^2)^2 \\
P_{2,1} &=& (1-2z)k_3,~~~
P_{2,2} = -(1-2z)k_2.
\eea
%

\subsection{The box coefficient $c_4^{0m}$}

The box coefficient $c_4^{0m}$ receives contributions from both $N_1$ and $N_2$,
and can be correspondingly decomposed as:
\bea
c_4^{0m} = - {2 \over (K^2)^2} C[Q_1,Q_2,K]^{(1)}
           + {\vev{1~2}\cb{3~4} \over K^2} C[Q_1,Q_2,K]^{(2)} \ .
\eea
We discuss the computation of $C[Q_1,Q_2,K]^{(1)}$ and $C[Q_1,Q_2,K]^{(2)}$
in detail, starting from the  expression given in Eq.(\ref{box-exp}).

\bigskip

\noindent $\bullet \ C[Q_1,Q_2,K]^{(1)}$

For the $N_1$ term, with $n=2$, the expression is given by
\bea C[Q_1,Q_2,K]^{(1)} & = & {(K^2)^{4}\over 2}\left({
\gb{P_{21,1}|R_1 |P_{21,2}}^2\gb{P_{21,1}|R_2 |P_{21,2}}^2
\over \gb{P_{21,1}|K|P_{21,2}}^4}+ {
\gb{P_{21,2}|R_1 |P_{21,1}}^2\gb{P_{21,2}|R_2 |P_{21,1}}^2
\over \gb{P_{21,2}|K|P_{21,1}}^4}
\right)
\eea

For analytic simplification, the following trace identity is helpful.
\bean \gb{P_1|R|P_2} \gb{P_2|S|P_1}  =  \Tr\left( {1-\gamma_5\over
2} \not{P_1} \not{R} \not{P_2} \not{S}\right)  \equiv \trm(P_1 R P_2
S) ~~~~~\Label{spinor-formula}\eean
In terms of vectors,
\bean
\trm(V_1 V_2 V_3 V_4) & = & {1\over 2} ( (2V_1 \cdot V_2)(2 V_3\cdot V_4)+(2V_1 \cdot V_4)(2 V_2\cdot V_3)
-(2V_1\cdot V_3)(2 V_2 \cdot V_4) -4i\eps_{\mu \nu\sigma \rho}V_1^\mu
V_2^\nu V_3^\sigma V_4^\rho)
\eean

The coefficient can then be expressed in terms of traces, and evaluated as follows.
\bea & &C[Q_1,Q_2,K]^{(1)}\nonumber \\ & & =   {(K^2)^4
  \left( (\trm(P_{21,2} K P_{21,1} R_1)\trm^2(P_{21,2} K P_{21,1}
R_2))^2 + (\trm(P_{21,1} K P_{21,2} R_1)\trm^2(P_{21,1} K P_{21,2}
R_2))^2 \right) \over 2 (\trm
(P_{21,1} K P_{21,2} K))^4 }\nonumber \\
& & =
 {(K^2)^4  z^2 (1-z)^2 \cb{3~4}^2 \vev{1~2}^2 \over   s_{12}^2 }
\eea

\bigskip

\noindent $\bullet \ C[Q_1,Q_2,K]^{(2)}$

For the $N_2$ term with $n=0$ the expression is given by

\bea C[Q_1,Q_2,K]^{(2)} & = & {(K^2)^{2}\over 2}\left({
\gb{P_{21,1}|R_1 |P_{21,2}}\gb{P_{21,1}|R_2 |P_{21,2}}
\over \gb{P_{21,1}|K
|P_{21,2}}^{2}}+ \{P_{21,1}\leftrightarrow P_{21,2}\}
\right).
\eea
Combining the two terms over a common denominator, we have
\bea
 & & C[Q_1,Q_2,K]^{(2)} \nonumber \\ & & =  {(K^2)^2
\left( \trm(P_{21,2} K P_{21,1} R_1)\trm(P_{21,2} K P_{21,1} R_2) +
\trm(P_{21,1} K P_{21,2} R_1)\trm(P_{21,1} K P_{21,2} R_2) \right)
\over 2 (\trm (P_{21,1} K P_{21,2} K))^2
}\nonumber \\
& & =
 {(K^2)^2  z (1-z) \cb{3~4}\vev{1~2} \over
 s_{12}
}
\eea

\bigskip

\noindent $\bullet \ $ The result of $c_4^{0m}$

We add our two contributions together and replace $z$ using (\ref{z-mu-ex2}).
The total box coefficient is thus
\bea
c_4^{0m}  & = & -{2 \over (K^2)^2} {(K^2)^4  z^2 (1-z)^2 \cb{3~4}^2 \vev{1~2}^2 \over
  s_{12}^2 }
+{\vev{1~2}\cb{3~4} \over K^2}
{(K^2)^2  z (1-z) \cb{3~4}\vev{1~2} \over   s_{12} } \\
 & = &
{(m^2+\mu^2) \cb{3~4}^2 \vev{1~2}^2 \over   s_{12} }
\left(
1 - {2(m^2+\mu^2) \over s_{12}}
\right)
\eea
%

\subsection{The triangle coefficients $c_{[23|4|1]}$ and $c_{[2|3|41]}$}

Both terms exhibit the symmetry of the amplitude, so our two triangles
are not independent.

The triangle coefficients $c_{[23|4|1]}$ and $c_{[2|3|41]}$
receive contributions from both $N_1$ and $N_2$,
and can be correspondingly decomposed as:
\bea
c_{[23|4|1]}& = & - {2 \over (K^2)^2} C[Q_1,K]^{(1)}
           + {\vev{1~2}\cb{3~4} \over K^2} C[Q_1,K]^{(2)} \\
c_{[2|3|41]}& = & - {2 \over (K^2)^2} C[Q_2,K]^{(1)}
           + {\vev{1~2}\cb{3~4} \over K^2} C[Q_2,K]^{(2)}
\eea

We discuss in parallel, first the contribution due to
$N_1$ to both coefficients, namely $C[Q_1,K]^{(1)}$ and $C[Q_2,K]^{(1)}$,
and later the one due to $N_2$, namely $C[Q_1,K]^{(2)}$ and $C[Q_2,K]^{(2)}$,
where the  triangle coefficient was given in Eq.(\ref{tri-exp}).

\bigskip

\noindent $\bullet \ C[Q_1,K]^{(1)}$ and $C[Q_2,K]^{(1)}$

Since the $N_1$ term with $n=2$, the triangle coefficient expression
is given by
\bea
& & C[Q_1,K]^{(1)}  =  { (K^2)^{3}\over
2}\frac{1}{(\sqrt{\Delta_1})^{3}}\frac{1}{3!
\vev{4~1}^{3}}
\frac{d^{3}}{d\tau^{3}}\left.\left({\prod_{j=1}^{4}
\vev{4-\tau 1 |R_j Q_1|4-\tau 1}\over
 \vev{4-\tau 1|Q_2 Q_1|4-\tau 1}}
+ {\prod_{j=1}^{4}
\vev{1-\tau 4 |R_j Q_1|1-\tau 4}\over
 \vev{1-\tau 4|Q_2 Q_1|1-\tau 4}}
\right)\right|_{\tau=0}\nonumber  \\
& & =  {(1-2z) \over 2 }\frac{1}{3! \vev{4~1}}
\frac{d^{3}}{d\tau^{3}}\left.\left({  \tgb{4|Q_1|4}^2 \vev{4-\tau
1, 2}^2 \tgb{3|Q_1|4-\tau 1}^2 \over
 \vev{4-\tau 1|Q_2 Q_1|4-\tau 1}}
+ {\tau^4
 \tgb{4|Q_1| 4}^2
\vev{1-\tau 4, 2}^2 \tgb{3|Q_1|1-\tau 4}^2
\over
 \vev{1-\tau 4|Q_2 Q_1|1-\tau 4}}
\right)\right|_{\tau=0}
 \nonumber \\
& & =  {(1-2z) (1-z)^2 (K^2)^2\over
12 \vev{4~1}}
\frac{d^{3}}{d\tau^{3}}\left.\left({
\vev{4-\tau 1, 2}^2 \tgb{3|Q_1|4-\tau 1}^2
\over
 \vev{4-\tau 1|Q_2 Q_1|4-\tau 1}}
\right)\right|_{\tau=0} \nonumber \\
& & =  { (1-z)^2 (K^2)^2 \over 12 }
\frac{d^{3}}{d\tau^{3}}\left.\left({ (\vev{4~2}-\tau\vev{1~2})^2
((1-z)\cb{3~1} + \tau z \cb{3~4})^2 \over -(1-z) \gb{4|3|1}+\tau
s_{34} +\tau^2 z \gb{1|3|4}} \right)\right|_{\tau=0} \nonumber \\
& & = 0. \eea
A similar calculation shows that
\bea
& & C[Q_2,K]^{(1)}  =  { (K^2)^{3}\over
2}\frac{1}{(\sqrt{\Delta_2})^{3}}\frac{1}{3!
\vev{3~2}^{3}}
 \frac{d^{3}}{d\tau^{3}}\left.\left({\prod_{j=1}^{4}
\vev{3-\tau 2 |R_j Q_2|3-\tau 2}\over
 \vev{3-\tau 2|Q_1 Q_2
|3-\tau 2}}
+ {\prod_{j=1}^{4}
\vev{2-\tau 3 |R_j Q_2|2-\tau 3}\over
 \vev{2-\tau 3|Q_1 Q_2
|2-\tau 3}}
\right)\right|_{\tau=0} \nonumber \\
& & = 0,
\eea
which can also be seen by the symmetry of the amplitude and the cut.

\bigskip

\noindent $\bullet \ C[Q_1,K]^{(2)}$ and $C[Q_2,K]^{(2)}$

For the $N_2$ term with $n=0$, the expression is simpler as
\bea
& & C[Q_1,K]^{(2)}  =  -{ (1-2z)(1-z) K^2 \over
2} \frac{d}{d\tau}\left.\left({
\vev{4-\tau 1~2}\tgb{3|Q_1|4-\tau
  1}
\over
 \vev{4-\tau 1|Q_2 Q_1|4-\tau 1}}
+ { \tau^2
\vev{1-\tau 4~2}\tgb{3|Q_1|1-\tau 4}\over
 \vev{1-\tau 4|Q_2 Q_1
|1-\tau 4}}
\right)\right|_{\tau=0} \nonumber \\
& & = -{ (1-2z)(1-z) K^2 \over
2} \frac{d}{d\tau}\left.\left({
(\vev{4~2}-\tau\vev{1~2})
(\tgb{3|Q_1|4} -\tau \tgb{3|Q_1|  1})
\over
 \vev{4-\tau 1|Q_2 Q_1|4-\tau 1}}
\right)\right|_{\tau=0}
\nonumber  \\
& & = 0. \eea
A similar calculation shows
\bea & & C[Q_2,K]^{(2)}    =  {(1-2z) \over 2 }
 \frac{d}{d\tau}\left.\left({
\vev{3-\tau 2~1}\tgb{4|Q_2|3-\tau 2}
\tgb{3|Q_2|3}
\over
\vev{3-\tau 2|Q_1 Q_2|3-\tau 2}}
+ { \tau^2 \vev{2-\tau 3~1}\tgb{4|Q_2|2-\tau 3}
\tgb{3|Q_2| 3}\over
 \vev{2-\tau 3|Q_1 Q_2|2-\tau 3}}
\right)\right|_{\tau=0}
\nonumber  \\
& & = 0, \eea
which can also be seen by symmetry.

\bigskip

\noindent {$\bullet \ $The results of $c_{[23|4|1]}$ and $c_{[2|3|41]}$}

Every term vanishes separately, so
\bea c_{[23|4|1]}& = & 0,~~~~ c_{[2|3|41]} = 0 .  \eea
The vanishing results for triangle coefficients
is not obvious from the beginning.
We suspect that there should be a more
directly physical argument to see this point.

\subsection{The bubble coefficient $c_{[23|41]}$}

The bubble coefficient $c_{[23|41]}$
receives contributions from both $N_1$ and $N_2$,
and can be correspondingly decomposed as:
\bea
c_{[23|41]}& = & - {2 \over (K^2)^2} C[K]^{(1)}
           + {\vev{1~2}\cb{3~4} \over K^2} C[K]^{(2)}
\eea

There is one subtlety regarding the calculation of the bubble coefficient.
The formulas involve an arbitrarily chosen, generic auxiliary null vector $\eta$. If  $\eta$ coincides with one of the $K_i$,
we need to use a modified formula, given in Appendix B.3.1 of \cite{Britto:2007tt}.  In this example, we illustrate both options.  First, we show the result with a generic choice of $\eta$; second, we use the formulas for the case $\eta=K_1$. Both are suitable for numerical evaluation, while the
special choice of $\eta$ may simplify the analytic expression.  We will find that the two results agree with each other, as well as with \cite{Bern:1995db}.

\subsubsection{Generic reference momentum $\eta$}

Let us start with  the formulas for generic $\eta$, given in
Eq.(\ref{bub-exp}). There are two terms we need to calculate.\\

\noindent $\bullet \ C[K]^{(1)}$

For the first term, with $N_1$ in the numerator, and $n=2$, the coefficient is
\bea
 C[K]^{(1)} = (K^2)^3 \sum_{q=0}^2 {(-1)^q\over q!} {d^q \over
ds^q}\left.\left( {\cal B}_{2,2-q}^{(0)}(s)+\sum_{r=1}^2\sum_{a=q}^2
\left({\cal B}_{2,2-a}^{(r;a-q;1)}(s)-{\cal
B}_{2,2-a}^{(r;a-q;2)}(s)\right)\right)\right|_{s=0},
\label{gggg:bubble-coeff}
\eea
where
\bea {\cal B}_{2,2-q}^{(0)}(s) =  {d^2\over d\tau^2}\left.\left(
{1 \over 2 [\eta|\W \eta K|\eta]^{2}}  {(2\eta\cdot
K)^{3-q} \over (3-q) (K^2)^{3-q}}{\prod_{j=1}^{4} \vev{\ell|R_j
(K+s\eta)|\ell}\over \vev{\ell~\eta}^{3} \prod_{p=1}^2 \vev{\ell|
Q_p(K+s\eta)|\ell}}|_{\ket{\ell}\to |K-\tau \W \eta|\eta]
}\right)\right|_{\tau= 0},
\eea
\bea & & {\cal B}_{2,2-a}^{(r;a-q;1)}(s)  =   {(-1)^{a-q+1}\over
 (a-q)! \sqrt{\Delta_r}^{a-q+1} \vev{P_{r,1}~P_{r,2}}^{a-q}}{d^{a-q} \over d\tau^{a-q}}
\left({1\over (3-a)} {\gb{P_{r,1}-\tau P_{r,2}|\eta|P_{r,1}}^{3-a}\over
\gb{P_{r,1}-\tau P_{r,2}|K|P_{r,1}}^{3-a}}\right. \nonumber \\ & & \times \left.\left.
{\vev{P_{r,1}-\tau P_{r,2}|Q_r \eta|P_{r,1}-\tau P_{r,2}}^{a-q} \prod_{j=1}^{4} \vev{P_{r,1}-\tau P_{r,2}|R_j
(K+s\eta)|P_{r,1}-\tau P_{r,2}}\over \vev{P_{r,1}-\tau P_{r,2}|\eta K|P_{r,1}-\tau P_{r,2}}^{3} \prod_{p=1,p\neq
r}^2 \vev{P_{r,1}-\tau P_{r,2}|
Q_p(K+s\eta)|P_{r,1}-\tau P_{r,2}}}\right)\right|_{\tau=0},
\eea
\bea & & {\cal B}_{2,2-a}^{(r;a-q;2)}(s)  = {(-1)^{a-q+1}\over
 (a-q)! \sqrt{\Delta_r}^{a-q+1} \vev{P_{r,1}~P_{r,2}}^{a-q}}{d^{a-q} \over d\tau^{a-q}}
\left({1\over (3-a)} {\gb{P_{r,2}-\tau P_{r,1}|\eta|P_{r,2}}^{3-a}\over
\gb{P_{r,2}-\tau P_{r,1}|K|P_{r,2}}^{3-a}}\right. \nonumber \\ & & \times \left.\left.
{\vev{P_{r,2}-\tau P_{r,1}|Q_r \eta|P_{r,2}-\tau P_{r,1}}^{a-q} \prod_{j=1}^{4} \vev{P_{r,2}-\tau P_{r,1}|R_j
(K+s\eta)|P_{r,2}-\tau P_{r,1}}\over \vev{P_{r,2}-\tau P_{r,1}|\eta K|P_{r,2}-\tau P_{r,1}}^{3} \prod_{p=1,p\neq
r}^2 \vev{P_{r,2}-\tau P_{r,1}|
Q_p(K+s\eta)|P_{r,2}-\tau P_{r,1}}}\right)\right|_{\tau=0}.
\eea
After making some substitutions, and considering the summation ranges of $a$ and $q$,  we get
\bea & & {\cal B}_{2,2-a}^{(1;a-q;1)}(s)  =  0,
\eea
\bea & & {\cal B}_{2,2-a}^{(1;a-q;2)}(s)  = {(-1)^{a-q+1}\over
 (a-q)! \sqrt{\Delta_1}^{a-q+1} \vev{4~1}^{a-q}}{d^{a-q} \over d\tau^{a-q}}
\left({1\over (3-a)} {\gb{1-\tau 4|\eta|1}^{3-a}\over
\gb{1-\tau 4|K|1}^{3-a}}\right. \nonumber \\ & & \times \left.\left.
{\vev{1-\tau 4|Q_1 \eta|1-\tau 4}^{a-q}
\vev{-\tau 4|R_1(K+s\eta)|1-\tau 4}^2
\vev{1-\tau 4|R_2(K+s\eta)|1-\tau 4}^2
\over \vev{1-\tau 4|\eta K|1-\tau 4}^{3}  \vev{1-\tau 4|
Q_2(K+s\eta)|1-\tau 4}}\right)\right|_{\tau=0}.
\eea
\bea & & {\cal B}_{2,2-a}^{(2;a-q;1)}(s)  =   {(-1)^{a-q+1}\over
 (a-q)! \sqrt{\Delta_2}^{a-q+1} \vev{3~2}^{a-q}}{d^{a-q} \over d\tau^{a-q}}
\left({1\over (3-a)} {\gb{3-\tau 2|\eta|3}^{3-a}\over
\gb{3-\tau 2|K|3}^{3-a}}\right. \nonumber \\ & & \times \left.\left.
{\vev{3-\tau 2|Q_2 \eta|3-\tau 2}^{a-q}
\vev{3-\tau 2|R_1(K+s\eta)|3-\tau 2}^2
\vev{3|R_2(K+s\eta)|3-\tau 2}^2
\over \vev{3-\tau 2|\eta K|3-\tau 2}^{3}  \vev{3-\tau 2|
Q_1(K+s\eta)|3-\tau 2}}\right)\right|_{\tau=0},
\eea
\bea & & {\cal B}_{2,2-a}^{(2;a-q;2)}(s)  = 0.
\eea

\bigskip

\noindent $\bullet \ C[K]^{(2)}$

For the second term $N_2$  with $n=0$,  the expression is much
simpler:
\bea
 C[K]^{(2)} = K^2 \left.\left( {\cal B}_{0,0}^{(0)}(s)+\sum_{r=1}^2
\left({\cal B}_{0,0}^{(r;0;1)}(s)-{\cal
B}_{0,0}^{(r;0;2)}(s)\right)\right)\right|_{s=0},
\eea
where
\bea {\cal B}_{0,0}^{(0)}(s=0) &=& \left.\left(
 {(2\eta\cdot
K) \over  K^2}{\vev{\ell|R_1 K|\ell}\vev{\ell|R_2 K|\ell} \over
\vev{\ell~\eta} \prod_{p=1}^2 \vev{\ell| Q_p
K|\ell}}\right)\right|_{\ket{\ell}\to |K|\eta] }
\nonumber \\
& = &{1 \over K^2}
{\cb{\eta|K R_1 |\eta}\cb{\eta|K R_2 |\eta}
\over  \cb{\eta|K Q_1 |\eta}\cb{\eta|K Q_2 |\eta}}
\nonumber  \\
 &=&
- {1 \over K^2}
{\cb{\eta~4} \cb{\eta~3}
\over  \cb{\eta~1}\cb{\eta~2} }
\\
  {\cal B}_{0,0}^{(1;0;1)}(s=0)  & = &     -   {1 \over
 \sqrt{\Delta_1} }
 {\gb{4|\eta|4}\over
\gb{4|K|4}} { \vev{4|R_1K|4} \vev{4|R_2K|4} \over \vev{4|\eta K|4}
\vev{4|Q_2 K|4}}
\nonumber \\
&=&       -  {1 \over K^2 }
{\cb{\eta~4} \vev{4~2}
\over \cb{\eta~ 1}  \vev{4~3}}
\\
 {\cal B}_{0,0}^{(1;0;2)}(s=0)
&=& 0.
\\
  {\cal B}_{0,0}^{(2;0;1)}(s=0)  & = &     -   {1 \over
 \sqrt{\Delta_2} }
 {\gb{3|\eta|3} \over \gb{3|K|3}}
{ \vev{3|R_1K|3} \vev{3|R_2K|3} \over \vev{3|\eta K|3} \vev{3|Q_1
    K|3}}
\nonumber \\
&=&
 {1 \over K^2  }
{\cb{\eta~3} \cb{4~2}  \over \cb{\eta~ 2} \cb{1~ 2}}
\\
{\cal B}_{0,0}^{(2;0;2)}(s=0)
&=& 0.
\eea

\bigskip

\noindent $\bullet$ Results:

 We
have used the numerical routines of {\tt S@M} \cite{Maitre:2007jq}
to show that while each single term $\cal{B}$ entering Eq.(\ref{gggg:bubble-coeff})
is $\eta$-dependent, their combination is
indeed independent of the choice of $\eta$.
The choice $\eta=k_3$ is found to be convenient. (Note that $k_3$ is not proportional to either of $K_1$, $K_2$.) Therefore we set $\eta=k_3$,
and we obtain the following analytic result.
\bea c_{[12|34]} &=& - {2 \over (K^2)^2} \bigg\{ { \vev{1~3}
\cb{1~3} (K^2)^{3}\over
  \cb{1~2}^2 \vev{4~3}^{2}} \left( {1 \over 2}-4z(1-z) \right)
- {5  (K^2)^{4} \over 3  \cb{1~2}^2 \vev{4~3}^{2}}z(1-z) + {
(K^2)^{4} \over 6  \cb{1~2}^2 \vev{4~3}^{2}} \nn & &
+
{K^2 \cb{1~3}\vev{1~3} \over    \vev{4~3}^2 \cb{1~2}^2}
\left( ({1 \over 6} - {2 \over 3}z(1-z)) \vev{1~3}^2 \cb{1~3}^2 +  (
{1\over 2}-3z(1-z))  K^2 \vev{1~3} \cb{1~3}
 \right)
  \bigg\}
+ {\vev{1~2}\cb{3~4} \over K^2}
  {\spb 4.3 \over \spb 1.2}
\nn &=& {\vev{1~2}^2 \cb{3~4}^2 \over 3s^2t^2} \left( -4
{(m^2+\mu^2) s } +6(m^2+\mu^2)t -2 st \right). \eea
We have used the relations $K^2=t, \vev{1~3}[1~3]=-s-t$.

\subsubsection{Special choice of  $\eta$}

Alternatively, we discuss the calculation of the bubble coefficient
by using the special choice of
$\eta=K_1=-k_1$  from the beginning.
With this choice, we need to use formulas for the $\cal{B}$ which are
slightly different from the ones used in the previous section.  They
are given in Appendix B.3.1 of \cite{Britto:2007tt}.  

Our convention for the spinors is:
\bea
\ket{\eta}=\ket{1}, \qquad |\eta ]=-|1].
\eea
\bea C[K]_n = (K^2)^{1+n} \sum_{q=0}^n {(-1)^q\over q!} {d^q \over
ds^q}\left.\left( {\cal B}_{n,n-q}^{(0)}(s)+\sum_{r=2}^k\sum_{a=q}^n
\left({\cal B}_{n,n-a}^{(r;a-q;1)}(s)-{\cal
B}_{n,n-a}^{(r;a-q;2)}(s)\right)\right)\right|_{s=0}.~~~~\Label{spe-Re-gen-n-1}\eea
\bea {\cal B}_{n,t}^{(0)}(s) &=& -{d^{n+1}\over
d\tau^{n+1}}\left.\left( { 1\over (1-2z) - s z }{[\eta|\W \eta
K|\eta]^{-n-1}\over (t+1) (n+1)!} {\prod_{j=1}^{n+k} \vev{\ell|R_j
(K+s\eta)|\ell}\over \vev{\ell~\eta}^{n+2} \prod_{p=2}^k \vev{\ell|
Q_p(K+s\eta)|\ell}}|_{\ket{\ell}\to |K-\tau \W \eta|\eta]
}\right)\right|_{\tau\to 0} \eea
Since $k=2$, we can directly set $r=2$:
\bea {\cal B}_{n,t}^{(2;a;1)}(s) & = & { 1\over (1-2z) - s z
}{(-1)^{a}\over (1-2z)^{a+1} (K^2)^{a+t+2}
 a!
\vev{3~2}^a}{d^a \over d\tau^a} \left({  \cb{1~3}^{t+1}\over (t+1)}
\right.
\nonumber \\ & & \times \left.\left.
{
\gb{\ell|Q_2|1}^a
\prod_{j=1}^{n+2} \vev{\ell|R_j (K+s\eta)|\ell}
\over
\vev{\ell~1}^{n+1-t-a} \cb{1~4}^{n+2} \vev{4~\ell}^{n+2}
}\right)
\right|_{\ket{\ell}=\ket{3}-\tau\ket{2}}
\eea
\bea {\cal B}_{n,t}^{(2;a;2)}(s) & = & { 1\over (1-2z) - s z
}{(-1)^{a}\over (1-2z)^{a+1} (K^2)^{a+t+2}
 a!
\vev{3~2}^a}{d^a \over d\tau^a} \left({ \cb{1~2}^{t+1} \over (t+1)}
\right.
\nonumber \\ & & \times \left.\left.
{
\gb{\ell|Q_2|1}^a
\prod_{j=1}^{n+2} \vev{\ell|R_j (K+s\eta)|\ell}
\over
\vev{\ell~1}^{n+1-t-a} \cb{1~4}^{n+2} \vev{4~\ell}^{n+2}
}\right)
\right|_{\ket{\ell}=\ket{2}-\tau\ket{3}}
\eea
Now we proceed to evaluate.\\

\noindent $\bullet \ C[K]^{(2)}$

For the $N_2$ term with $n=0$, the evaluation is simple.  In particular, there are no derivatives in $s$, so we can set $s=0$ directly.
\bea C[K]^{(2)} = K^2 \left.\left( {\cal B}_{0,0}^{(0)}(s)+
{\cal B}_{0,0}^{(2;0;1)}(s)-{\cal
B}_{0,0}^{(2;0;2)}(s)\right)\right|_{s=0}.
\eea
%
%
%
Choosing $\W\eta=3$, we have
\bea {\cal B}_{0,0}^{(0)}(s=0) & = &
{1 \over \cb{1~2} \vev{3~4}},~~~~ {\cal B}_{0,0}^{(2;0;1)}(s=0) = {1
\over K^2}{\cb{1~3}\cb{4~2}\over \cb{1~2}^{2} },~~~{\cal
B}_{0,0}^{(2;0;2)}(s=0)=0\eea
so the total contribution comes to
\bea C[K]^{(2)} =
{\cb{4~3} \over \cb{1~2}}
\eea

\bigskip

\noindent $\bullet \ C[K]^{(1)}$

For the $N_1$ term with $n=2$, the calculation is a bit more involved.

\bea C[K]^{(1)} = (K^2)^{3} \sum_{q=0}^2 {(-1)^q\over q!} {d^q \over
ds^q}\left.\left( {\cal B}_{2,2-q}^{(0)}(s)+ \sum_{a=q}^2
\left({\cal B}_{2,2-a}^{(2;a-q;1)}(s)-{\cal
B}_{2,2-a}^{(2;a-q;2)}(s)\right)\right)\right|_{s=0}.
\eea
For the various terms, we have
\bea
 {\cal B}_{2,2-q}^{(0)}(s) &=& {1 \over (3-q) 3!
 \vev{3~4}^2 \cb{1~2}^2 }
  {d^{3}\over d\tau^{3}}\left(
{ (1-2z)^4 (1+s)^2  \over (1-2z) - s z }
  \times \right.~~~\Label{exm-2-1}
\\ & &
 \left. \left.
{ (1-\tau)^2
 \left( K^2 (1+s) - \tau (K^2 +s\cb{3~1}\vev{1~3})\right)^2
\over
\left( (1+s)(1-2z)K^2 + \tau ( (zs+2z-s-1)K^2 - s(1-2z)\cb{3~1}\vev{1~3})
+ \tau^2 s (1-z)\cb{3~1}\vev{1~3} \right)
}
\right)\right|_{\tau\to 0} \nonumber
\eea
%

%
%
\bea  {\cal B}_{2,2-a}^{(2;a-q;1)}(s) 
  & = &  {(1+s)^2 (1-2z)^{3-a+q} \over (1-2z) - s z
}{(-1)^{a-1} \vev{3~2}^{2} \cb{1~3}^{3-a}\over
  (K^2)^{4-q} (a-q)!
}  ~~~\Label{exm-2-2}\\ & & \times {d^{a-q} \over d\tau^{a-q}}
\left( { ((1-z)\cb{1~2}+ \tau z \cb{1~3})^{a-q}
(\vev{1~3}-\tau\vev{1~2})^{3-q} (K^2+s\cb{3~1}\vev{1~3}-s\tau
\cb{3~1}\vev{1~2})^2 \over
 (3-a) \cb{1~4}^2
(\vev{4~3}-\tau\vev{4~2})^{4} }\right)\nonumber \eea
The term $ {\cal B}_{2,2-a}^{(2;a-q;2)}(s)$ vanishes after taking the derivatives with respect to $s$.  The reason is the following.
 Notice
that
\bea & & {\cal B}_{n,t}^{(2;a-q;2)}(s)
\\
 & & = {(1+s)^2  (1-2z)^{3-a+q} \over (1-2z) - s z
}{(-1)^{a-q} \vev{3~2}^2 \cb{1~2}^{t+1} \over (K^2)^{a-q+t+2}
 (a-q)!
\vev{3~2}^{a-q}}{d^{a-q} \over d\tau^{a-q}} \left(
{  \tau^2 \over (t+1)}
\left.
{
\gb{\ell|Q_2|1}^{a-q}
 \tgb{3|K+s\eta|\ell}^2
\over
\vev{\ell~1}^{-1-t-a+q} \cb{1~4}^{2} \vev{4~\ell}^{4}
}\right)
\right|_{\ket{\ell}=\ket{2}-\tau\ket{3}}. \nonumber
\eea
We can see that the $\tau$-derivative vanishes unless $a-q=2$, in which case we get
\bea {\cal B}_{n,t}^{(2;2;2)}(s)
 & = & {(1+s)^2  (1-2z) \over (1-2z) - s z
}{         \vev{3~2}^2 \cb{1~2}^{t+1} \over (K^2)^{4+t}
 \vev{3~2}^2} \left(
{  1\over (t+1)}
{
(-z)^2  \gb{2|3|1}^2
s^2  \tgb{3|1|2}^2
\over
\vev{2~1}^{-3-t} \cb{1~4}^{2} \vev{4~2}^{4}
}\right)
\eea
However, the condition $a-q=2$ implies $a=2,q=0$.  Therefore we can set $s=0$, and the expression vanishes:
\bea {\cal B}_{n,t}^{(2;2;2)}(s) =0.~~~\Label{exm-2-3} \eea
Now we collect the results of (\ref{exm-2-1}),(\ref{exm-2-2}) and
(\ref{exm-2-3}). We take $\W\eta=k_3$.
Define
\bea
C_1 \equiv  {\cb{1~3}\vev{1~3} \over (K^2)^{2}
\cb{1~2}^2 \vev{4~3}^2 }.
\eea
Let us begin with the terms with $q=0$:
\bea {\cal B}_{2,2}^{(0)}(s=0) &=& - {K^2(1-2z)^2\over
3\vev{3~4}^2\cb{2~1}^2 } \\{\cal B}_{2,2}^{(2;0;1)}(s=0)
 & = & C_1
 (-{1\over 3})(1-2z)^2  \cb{1~3}^{2} \vev{1~3}^{2}\nonumber \\ {\cal B}_{2,1}^{(2;1;1)}(s=0)
 & = &
C_1 \left( -{1\over 2}(1-2z)^2  \cb{1~3}^2 \vev{1~3}^2 + ({3\over
2}-{9\over 2}z+3z^2)  K^2\cb{1~3}\vev{1~3}
 \right) \nonumber \\ {\cal B}_{2,0}^{(2;2;1)}(s=0)
& = & C_1 \left( (-3+6z-3z^2) (K^2)^2 -(1-2z)^2
\vev{1~3}^2\cb{1~3}^2 +(6-18z+12z^2)K^2\cb{1~3} \vev{1~3} \right)
\nonumber\eea
For $q=1$:
\bea - \left. {d \over ds} {\cal B}_{2,1}^{(0)}(s) \right|_{ s = 0}
&=& - {(1-2z)^2  \over 2
 \vev{3~4}^2 \cb{1~2}^2 }
{ \cb{3~1}\vev{1~3} - (1-2z) K^2 \over (1-2z)}
\\ - \left. {d \over ds}{\cal B}_{2,1}^{(2;0;1)}(s)\right|_{ s
= 0}
 & = & C_1
 \left( (1-2z)^2 \cb{1~3}^2\vev{1~3}^2
+  (-1+{7 \over 2} z-3z^2) K^2 \vev{1~3} \cb{1~3}\right) \nonumber
\\  - \left.{d \over ds} {\cal B}_{2,0}^{(2;1;1)}(s)\right|_{s=0}
  &= & C_1
\left( (4-10z+6z^2)(K^2)^2+ 2(1-2z)^2 \cb{1~3}^2\vev{1~3}^2 +
(-10+30z-21 z^2)K^2 \cb{1~3}\vev{1~3} \right) \nonumber \eea
For $q=2$:
\bea \left. {1 \over 2}{d^2 \over ds^2}
 {\cal B}_{2,0}^{(0)}(s) \right|_{s=0}
 &=& { 1 \over
 \vev{3~4}^2 \cb{1~2}^2 }
{ z \left( (1-2z)\cb{3~1}\vev{1~3} -(1-z)K^2 \right)
 }\\ \left. {1 \over 2}{d^2 \over ds^2}  {\cal
    B}_{2,0}^{(2;0;1)}(s)\right|_{s=0} &= & C_1
\left(
 (-1+2z-z^2) (K^2)^2  -(1-2z)^2 \vev{1~3}^2 \cb{1~3}^2
+ (4 - 14z + 12z^2) K^2 \vev{1~3} \cb{1~3} \right) \nonumber \eea

All together, we get the following result for the $N_1$ term.
\bea C[K]^{(1)} &=& { \vev{1~3} \cb{1~3} (K^2)^{3}\over
  \cb{1~2}^2 \vev{4~3}^{2}} \left( {1 \over 2}-4z(1-z) \right)
- {5  (K^2)^{4} \over 3  \cb{1~2}^2 \vev{4~3}^{2}}z(1-z) + {
(K^2)^{4} \over 6  \cb{1~2}^2 \vev{4~3}^{2}}
\\
& & +  {K^2 \cb{1~3}\vev{1~3} \over    \vev{4~3}^2 \cb{1~2}^2}
 \left( ({1 \over 6} - {2 \over 3}z(1-z)) \vev{1~3}^2
\cb{1~3}^2 +  ( {1\over 2}-3z(1-z))  K^2 \vev{1~3} \cb{1~3}
 \right)
\nonumber \eea

\bigskip

\noindent $\bullet \ $ The result of $c_{[12|34]}$

Final bubble coefficient:
\bea
c_{[12|34]} &=&
- {2 \over (K^2)^2} \bigg\{
{ \vev{1~3} \cb{1~3} (K^2)^{3}\over
  \cb{1~2}^2 \vev{4~3}^{2}} \left( {1 \over 2}-4z(1-z) \right)
- {5  (K^2)^{4} \over
3  \cb{1~2}^2 \vev{4~3}^{2}}z(1-z)
+
{  (K^2)^{4} \over
6  \cb{1~2}^2 \vev{4~3}^{2}}
\nn & & \qquad \qquad
+  {K^2
\cb{1~3}\vev{1~3} \over    \vev{4~3}^2 \cb{1~2}^2}
\left(
({1 \over 6} - {2 \over 3}z(1-z))
\vev{1~3}^2 \cb{1~3}^2
+  ( {1\over 2}-3z(1-z))  K^2 \vev{1~3} \cb{1~3}
 \right)
  \bigg\}
\nn
& &
+ {\vev{1~2}\cb{3~4} \over K^2}
  {\spb 4.3 \over \spb 1.2}
\nn
&=&
{\vev{1~2}^2 \cb{3~4}^2 \over 3s^2t^2}
\left( -4 {(m^2+\mu^2) s } +6(m^2+\mu^2)t -2 st \right)
\eea
where we used the definitions $K^2=t, \vev{1~3}[1~3]=-s-t$.

\subsection{Comparison with the literature}

The $t$-channel cut of $A(1^-,2^-,3^+,4^+)$ admits a decomposition
in terms of cuts of master integrals as shown in Fig.\ref{fig:s23cut}.
Its expression was given in Eq.(5.33) of \cite{Bern:1995db}, and reads
\bea
\left. A^{\rm fermion}_4(1^-,2^-,3^+,4^+) \right|_{t-{\rm cut}} &=&
-2 \left. A^{\rm scalar}_4(1^-,2^-,3^+,4^+) \right|_{t-{\rm cut}}
- \left. {1 \over (4\pi)^{2-\eps}} A_4^{\rm tree} (t J_4-I_2(t))\right|_{t-{\rm cut}}
\eea
with
\bea
 \left. A^{\rm scalar}_4(1^-,2^-,3^+,4^+) \right|_{t-{\rm cut}} &=&
\left.  {1 \over (4\pi)^{2-\eps}} A_4^{\rm tree}
\left( {1 \over t} I_2^{(1,3),D=6-2\eps} + {1\over s}J_2^{(1,3)}
-{t \over s} K_4 \right) \right|_{t-{\rm cut}}
\eea
and
\bea
  A^{\rm tree}_4 &=&
i{\vev{1~2}^4 \over    \vev{1~2}\vev{2~3}\vev{3~4}\vev{4~1}} =
-i {\vev{1~2}^2 \cb{3~4}^2    \over K^2 s_{12}}
\eea
where we neglected the cut-free term, $I_1$ and $I_2(0)$.
In standard notation we have $s=s_{12}, \ t=s_{23}, \ u= - s - t = s_{13}$.

Now we translate the expression of \cite{Bern:1995db} into our canonical basis, using the identities of Appendix \ref{sec:IntegralsTranslation}.
\bea
\left. A^{\rm fermion}_4(1^-,2^-,3^+,4^+) \right|_{t-{\rm cut}}
&=&
-2 \left. A^{\rm scalar}_4(1^-,2^-,3^+,4^+) \right|_{t-{\rm cut}}
- \left. {1 \over (4\pi)^{2-\eps}} A_4^{\rm tree} (t
  J_4-I_2(t))\right|_{t-{\rm cut}} \\
&=&
- \left.  {1 \over (4\pi)^{2-\eps}} A_4^{\rm tree}
\left( {2 \over t} I_2^{(1,3),D=6-2\eps} + {2\over s}J_2^{(1,3)}
-{2t \over s} K_4
+ t  J_4-I_2(t)
\right) \right|_{t-{\rm cut}} \\
&=&
- i \left.  A_4^{\rm tree}
\left( {2 \over t}
\left( {t \over 6} I^{\rm BFM}_2[1] +{1 \over 3}I_1
-{2 \over 3}I^{\rm BFM}_2[m^2+\mu^2]
\right)
+ {2\over s} I^{\rm BFM}_2[m^2+\mu^2]
\nonumber \right. \right. \\ & & \left. \left.
-{2t \over s} I^{\rm BFM}_4[(m^2+\mu^2)^2]
+ t  I^{\rm BFM}_4[m^2+\mu^2]
- I^{\rm BFM}_2[1]
\right) \right|_{t-{\rm cut}} \\
&=&
  \left. {\vev{1~2}^2 \cb{3~4}^2 \over st}
\left(
 {2 \over 3} I^{\rm BFM}_2[1]
+{4 \over 3t}I^{\rm BFM}_2[m^2+\mu^2]
- {2\over s} I^{\rm BFM}_2[m^2+\mu^2]
\nonumber \right. \right. \\ & & \left. \left.
+{2t \over s} I^{\rm BFM}_4[(m^2+\mu^2)^2]
- t  I^{\rm BFM}_4[m^2+\mu^2]
\right) \right|_{t-{\rm cut}}
\eea
We have reproduced every one of these coefficients, up to an overall minus sign in the amplitude.

\section{From polynomials in $u$ to final coefficients}
 
As proven in the Appendix \ref{polynomialproof}, the coefficients of
2-, 3-, and 4-point functions in  four dimensions are polynomials
in $u$ (or equivalently $\mu^2$), of known degree $d$: for boxes, 
 $d= [(n+2)/2]$; for triangles,  $d=[(n+1)/2]$; for
bubbles,  $d=[n/2]$; where $[x]$ denotes the greatest integer less
than or equal to $x$.  Using this fact, we can generally represent
any coefficient of the master integral as 
\bea P_d(u) = \sum_{r=0}^d \ c_r \ u^r \ . \eea
The coefficients $c_r$ are in one-to-one correspondence to the
coefficients of the {\it shifted-dimension} master integrals
(see Section 2). \\
To compute the $c_r$ analytically, one can proceed with the standard
differentiations with respect to $u$, at $u=0$: 
\bea c_r & = &
\left. {1 \over k ! } {d^{r} \over d u^r} P_d(u) \right|_{u=0}.
\eea
When the differentiations are time consuming, or the analytic
expression is not needed, one can switch to the following numerical
procedure, and extract the $c_r$ algebraically, by {\it
projections}.
\begin{enumerate}
\item
Generate the values $P_{d,k}, (k=0,...,d-1)$, \bea P_{d,k} =
P_d(u_k) \ , \eea by evaluating $P_d(u)$ at particular points:
\bea
 u_k = \ e^{-2 \pi i {k / d}} \ .
\eea
\item
Using the orthogonality relations for plane waves, one can
obtain the coefficient $c_r$ simply by the following formula: 
\bea c_r &=& {1 \over d}
         \sum_{k=0}^{d-1} \ P_{d,k} \  e^{2 \pi i r {k /d} }.
\eea
\end{enumerate}

\acknowledgments

RB is supported by Stichting FOM. BF  is
 supported by Qiu-Shi Professor Fellowship from Zhejiang University,
 China.

\appendix

\section{Change of Basis}
\label{sec:IntegralsTranslation}

To compare our results to the literature, we need to convert the master integrals used in \cite{Bern:1995db, Rozowsky:1997dm} to our canonical $(4-2\eps)$-dimensional basis, (\ref{n-scalar}).
  For clarity, we now
 denote the basis used in this paper by $I^{\rm BFM}_n$, while the other integrals in this appendix are defined according to \cite{Bern:1995db, Rozowsky:1997dm}.  The first point is
then that
\bea
I_n = i (-1)^n  (4\pi)^{2-\eps} I^{\rm BFM}_n.
\eea
We use the identities from Appendix A.4 of \cite{Bern:1995db} to perform the conversion.
\bea
J_4 &=& I_4[m^2+\mu^2] = i  (4\pi)^{2-\eps} I^{\rm BFM}_4[m^2+\mu^2]  \\
I_2(t) &=& i  (4\pi)^{2-\eps} I^{\rm BFM}_2[1] \\
I_2^{(1,3),D=6-2\eps} &=& {t \over 6}I_2(t)+{1 \over 3}I_1-{2 \over 3}
 J_2^{(1,3)}
= i(4\pi)^{2-\eps} \left( {t \over 6} I^{\rm BFM}_2[1] +{1 \over 3}I_1
-{2 \over 3}I^{\rm BFM}_2[m^2+\mu^2]
\right)
\\
J_2^{(1,3)} &=&  J_2^{(1,3)}[m^2+\mu^2]
= i  (4\pi)^{2-\eps} I^{\rm BFM}_2[m^2+\mu^2] \\
K_4 &=& I_4[(m^2+\mu^2)^2]= i  (4\pi)^{2-\eps} I^{\rm BFM}_4[(m^2+\mu^2)^2]
\eea
%

\section{\label{polynomialproof}The $u$-dependence of the coefficients}

Here we analyze the $u$-dependence of the integral coefficients given by our formulas.  First, we prove that they are polynomials in $u$.  Then, we present some alternate formulas where this polynomial dependence is more explicit.  This material is a straightforward generalization of the analysis in the massless case \cite{bfy}, so we omit many of the details here.

To begin, we rewrite our vectors $R_j,Q_j$ from (\ref{R-def}) and (\ref{Q-def}) in the following way:
\bea R_j &=& -\left((1-2z)+{M_1^2-M_2^2\over K^2} \right)p_j+ \b_j K, \nonumber \\
& & p_j \equiv \left(P_j-{P_j\cdot
K\over K^2}K \right),  \\
& & \b_j \equiv -{(P_j\cdot
K)\over K^2}\left(1+{M_1^2-M_2^2\over K^2} \right). ~~~\Label{beta-mass} \\
Q_j &=& -\left((1-2z)+{M_1^2-M_2^2\over K^2} \right) q_j+ \a_j K,\\
& & q_j \equiv \left(K_j-{K_j\cdot
K\over K^2}K \right),   \\
& & \a_j  \equiv  -{(K_j\cdot
K)\over K^2}\left(1+{M_1^2-M_2^2\over K^2} \right)+
{K_j^2+M_1^2-m_j^2\over K^2}.~~~\Label{alpha-mass}\eea
Notice that
\bea
p_j \cdot K = 0, ~~~~~q_j \cdot K=0.
\eea
Using (\ref{z-rel-1}), we have
\bea R_j(u)=-\b(\sqrt{1-u}) p_j +\b_j K,~~~Q_j(u)=-\b (\sqrt{1-u}) q_j+\a_j
K.~~~\Label{R-Q-massive}\eea
This is the same expression as in the massless case, except for the factor $\b$. The point is that now all $u$-dependence
is in the factor $\sqrt{1-u}$, just as in the massless case.
In fact, we should now consider the factor $\b\sqrt{1-u}$ as our basic quantity.  The proof that the integral coefficients are polynomials in $u$ was performed by considering the (demonstrably finite) series expansion in $\sqrt{1-u}$, and showing that the odd powers drop out.  Therefore, the same arguments now carry over to the series expansion in $\b\sqrt{1-u}$.

\subsection{Triangle coefficients}

Let us begin with triangle coefficients.
The null vectors $P_{s,i}$ exhibit a simple dependence on $u$.  Specifically,
\bean
P_{s,i}(u)& = &
-\b(\sqrt{1-u}) P_{q_s,i}, \eean
where
\bea
P_{q_s,i} \equiv q_s \pm \left({\sqrt{-q_s^2 \over K^2}} \right)K,
~~~\Label{ufree-null}
\eea
which is manifestly independent of $u$.
In defining the spinor components of $P_{s,i}$, we can place the $u$-dependent factor inside the antiholomorphic spinor, i.e.,
\bea
\ket{P_{s,i}}=\ket{P_{q_s,i}},~~~~~~|P_{s_i}] = -\b(\sqrt{1-u}) |P_{q_s,i}].
~~~\Label{ufree-spinor}
\eea
Then, for the triangle coefficients, we have
\bea C[Q_s,K] & = & { (K^2)^{1+n}\over
2}\frac{1}{(-\b\sqrt{1-u})^{n+1}(\sqrt{-4q_s^2 K^2})^{n+1}}\frac{1}{(n+1)!
\vev{P_{q_s,1}~P_{q_s,2}}^{n+1}} \nonumber
\\ & & \times \frac{d^{n+1}}{d\tau^{n+1}}\left.\left({\prod_{j=1}^{k+n}
\vev{P_{q_s,1}-\tau P_{q_s,2} |R_j(u) Q_s(u)|P_{q_s,1}-\tau P_{q_s,2}}\over
\prod_{t=1,t\neq s}^k \vev{P_{q_s,1}-\tau P_{q_s,2}|Q_t(u) Q_s(u)
|P_{q_s,1}-\tau P_{q_s,2}}} + \{P_{q_s,1}\leftrightarrow
P_{q_s,2}\}\right)\right|_{\tau=0}.~~~~~\Label{tri-exp-0}\eea
Further, we make use of some identities,
\bean \vev{\ell| Q_t(u) Q_s(u)|\ell} &= &
\vev{\ell|(Q_t(u)-{\a_t\over \a_s} Q_s(u)) Q_s(u)|\ell}=
-\b\sqrt{1-u}\vev{\ell|(q_t-{\a_t\over \a_s} q_s) Q_s(u)|\ell}\\
\vev{\ell| R_t(u) Q_s(u)|\ell} &= &
-\b\sqrt{1-u}\vev{\ell|(p_t-{\b_t\over \a_s} q_s) Q_s(u)|\ell}\eean
along with the definitions
\bea \W q_t \equiv (q_t-{\a_t\over \a_s} q_s),~~~~\W p_t \equiv (p_t-{\b_t\over
\a_s} q_s).\eea
Our final form for the triangle coefficient is:
\bea C[Q_s,K] & = & { 1\over
2}
\left(\sqrt{K^2 \over -4q_s^2} \right)^{n+1}
\frac{1}{(n+1)!
\vev{P_{s,1}~P_{s,2}}^{n+1}} \nonumber
\\ & & \times \frac{d^{n+1}}{d\tau^{n+1}}\left.\left({\prod_{j=1}^{k+n}
\vev{P_{s,1}-\tau P_{s,2} |\W p_j Q_s(u)|P_{s,1}-\tau P_{s,2}}\over
\prod_{t=1,t\neq s}^k \vev{P_{s,1}-\tau P_{s,2}|\W q_t Q_s(u)
|P_{s,1}-\tau P_{s,2}}} + \{P_{s,1}\leftrightarrow
P_{s,2}\}\right)\right|_{\tau=0}.~~~~~\Label{tri-exp-1}\eea
Here, the $u$-dependence is concentrated entirely within the vector
$Q_s(u)$, since we have made sure to choose the spinor components
wisely in (\ref{ufree-spinor}), so that the holomorphic spinors are $u$-independent.

\subsection{Bubble coefficients}

We follow the same procedure as with triangles, and make use of the same definitions (\ref{ufree-null}), (\ref{ufree-spinor}).  The $u$-dependence can be concentrated within the vectors $R_j(u)$ and $Q_r(u),Q_p(u)$, along with the explicit factor of $\sqrt{1-u}$, in the following formulas.
\bea
 C[K] = (K^2)^{1+n} \sum_{q=0}^n {1\over q!} {d^q \over
ds^q}\left.\left( {\cal B}_{n,n-q}^{(0)}(s)+\sum_{r=1}^k\sum_{a=q}^n
\left({\cal B}_{n,n-a}^{(r;a-q;1)}(s)-{\cal
B}_{n,n-a}^{(r;a-q;2)}(s)\right)\right)\right|_{s=0},~~~~~\Label{bub-exp--1}
\eea
where
\bea {\cal B}_{n,t}^{(0)}(s)\equiv {d^n\over d\tau^n}\left.\left(\left. {1
\over n! [\eta|\W \eta K|\eta]^{n}}  {(2\eta\cdot K)^{t+1} \over
(t+1) (K^2)^{t+1}}{\prod_{j=1}^{n+k} \vev{\ell|R_j(u)
(K-s\eta)|\ell}\over \vev{\ell~\eta}^{n+1} \prod_{p=1}^k \vev{\ell|
Q_p(u)(K-s\eta)|\ell}}\right|_{\ket{\ell}\to |K-\tau \W \eta|\eta]
}\right)\right|_{\tau= 0},~~~\eea
\bea & & {\cal B}_{n,t}^{(r;b;1)}(s)  \equiv  {(-1)^{b+1}\over
 b! (-\b\sqrt{1-u})^{b+1}\sqrt{-4q_r^2 K^2}^{b+1} \vev{P_{r,1}~P_{r,2}}^b}{d^b \over d\tau^{b}}
\left({1\over (t+1)} {\gb{P_{r,1}-\tau
P_{r,2}|\eta|P_{q_r,1}}^{t+1}\over \gb{P_{r,1}-\tau
P_{_r,2}|K|P_{q_r,1}}^{t+1}}\right. \nonumber \\ & & \times
\left.\left. {\vev{P_{r,1}-\tau P_{r,2}|Q_r(u) \eta|P_{r,1}-\tau
P_{r,2}}^{b} \prod_{j=1}^{n+k} \vev{P_{r,1}-\tau P_{r,2}|R_j(u)
(K-s\eta)|P_{r,1}-\tau P_{r,2}}\over \vev{P_{r,1}-\tau P_{r,2}|\eta
K|P_{r,1}-\tau P_{r,2}}^{n+1} \prod_{p=1,p\neq r}^k
\vev{P_{r,1}-\tau P_{r,2}| Q_p(u)(K-s\eta)|P_{r,1}-\tau
P_{r,2}}}\right)\right|_{\tau=0},~~~\eea
\bea & & {\cal B}_{n,t}^{(r;b;2)}(s)  \equiv  {(-1)^{b+1}\over
 b! (-\b\sqrt{1-u})^{b+1}\sqrt{-4q_r^2 K^2}^{b+1} \vev{P_{r,1}~P_{r,2}}^{b}}{d^{b} \over d\tau^{b}}
\left({1\over (t+1)} {\gb{P_{r,2}-\tau
P_{r,1}|\eta|P_{q_r,2}}^{t+1}\over \gb{P_{r,2}-\tau
P_{r,1}|K|P_{q_r,2}}^{t+1}}\right. \nonumber \\ & & \times
\left.\left. {\vev{P_{r,2}-\tau P_{r,1}|Q_r(u) \eta|P_{r,2}-\tau
P_{r,1}}^{b} \prod_{j=1}^{n+k} \vev{P_{r,2}-\tau P_{r,1}|R_j(u)
(K-s\eta)|P_{r,2}-\tau P_{r,1}}\over \vev{P_{r,2}-\tau P_{r,1}|\eta
K|P_{r,2}-\tau P_{r,1}}^{n+1} \prod_{p=1,p\neq r}^k
\vev{P_{r,2}-\tau P_{r,1}| Q_p(u)(K-s\eta)|P_{r,2}-\tau
P_{r,1}}}\right)\right|_{\tau=0}.~~~\eea
%

\subsection{Box and pentagon coefficients}

Although the formula (\ref{box-exp}) for box and pentagon coefficients looks simple, the $u$-dependence now gets complicated.
We consider the separate cases $k=2$, $k=3$, and $k \geq 4$.

\subsubsection{The case  $k=2$}

In this case, there is only one box, and no pentagons.  The box coefficient is given by
\bea{(K^2)^{2+n}\over 2} \left( {\prod_{j=1}^{n+2} \gb{P_{(Q_j(u),
Q_i(u));1}(u)|R_j(u)|P_{(Q_j(u), Q_i(u));2}(u)}\over \gb{P_{(Q_j(u),
Q_i(u));1}(u)|K|P_{(Q_j(u), Q_i(u));2}(u)}^{n+2}}+\{ P_{(Q_j(u),
Q_i(u));1}(u)\leftrightarrow P_{(Q_j(u), Q_i(u));2}(u)\}\right) \eea

Given the vectors $Q_i,Q_j, K$ that select a particular box,
it is useful to construct a vector $q_0^{(q_i,q_j,K)}$ that is
orthogonal to all three, and independent of $u$:
\bea (q_0)_\mu^{(q_i,q_j,K)} &\equiv&
{1 \over K^2} \epsilon_{\mu\nu \rho \xi}
q_i^\nu q_j^\rho K^\xi \\
&=&{1 \over K^2}
 \epsilon_{\mu\nu \rho \xi} K_i^\nu K_j^\rho
K^\xi,~~~\Label{q-0}
\eea
As in the massless case, the $u$-dependence can be concentrated in a
single factor, $\a^{(q_i,q_j)}(u)$.  If all input quantities are set
to their values with $u=0$, except for adjusting the definition of $R_s(u)$
as follows,
\bea
& & R_s(u) \to \W R_s(u)  \equiv {p_s\cdot q_0^{(q_i,q_j,K)}\over
(q_0^{(q_i,q_j,K)})^2} (\a^{(q_i,q_j)}(u)-1)(-\b
q_0^{(q_i,q_j,K)})+R_s(u=0) \\
& & \a^{(q_i,q_j)}(u) \equiv {\sqrt{\b^2(1-u)+ {4K^2[
\a_i\a_j(2q_i\cdot q_j)-\a_i^2 q_j^2-\a_j^2 q_i^2]\over (2q_i\cdot
q_j)^2-4 q_i^2 q_j^2}}\over \sqrt{\b^2+ {4K^2[ \a_i\a_j(2q_i\cdot
q_j)-\a_i^2 q_j^2-\a_j^2 q_i^2]\over (2q_i\cdot q_j)^2-4 q_i^2
q_j^2}}},
~~~\Label{def-alpha-box-k2}
\eea
then the value of the box coefficient remains the same.

In summary, the box coefficient for $k=2$ is given by
\bea C[K_i,K_j]_{k=2} & = & {(K^2)^{2+n}\over 2} \left(
{\prod_{s=1}^{n+2} \gb{P_{ji;1}|\W R_s(u)|P_{ji;2}}\over \gb{P_{ji;1}|K|P_{ji;2}}^{n+2}}+\{
P_{ji;1}\leftrightarrow P_{ji;2}\}\right)~~\Label{k=2-final} \eea
where
\bea \W R_s(u) ={p_s\cdot q_0^{(q_i,q_j,K)}\over
(q_0^{(q_i,q_j,K)})^2} (\a^{(q_i,q_j)}(u)-1)(-\b
q_0^{(q_i,q_j,K)})+R_s(u=0)~~~\Label{W-R-S} \eea
and
\bea
P_{ji;a} = P_{(Q_j, Q_i);a}(u=0)
~~~\Label{pjia}
\eea
In evaluating (\ref{W-R-S}), it is useful to observe the following:
\bea { (p_s\cdot q_0^{(q_i,q_j,K)})}={\eps(p_s,q_i,q_j,K)\over
K^2}={\eps(P_s,K_i,K_j,K)\over
K^2}~~~\Label{a-0-k=2}\eea
 This formula (\ref{k=2-final})
looks the same as in the massless case; the difference is the
appearance of $\b$ in (\ref{W-R-S}), both explicitly and through the
definition (\ref{def-alpha-box-k2}) of $\a^{(q_i,q_j)}$.

\subsubsection{The case $k=3$}

Here there is a pentagon, as well as three boxes.
The differences from the massless case are all based in the
definitions of $R_j(u),Q_j(u)$:  there is always a factor of $\b$
accompanying $\sqrt{1-u}$, and mass parameters enter into the
definitions (\ref{beta-mass}), (\ref{alpha-mass}) of $\b_j,\a_j$.

When we make these adjustments, we find that the pentagon coefficient
takes the same form as in the massless case,
\bea C[Q_i, Q_j, Q_t] = (K^2)^{3+n} \prod_{s=1}^{n+3}
\b_s^{(q_i,q_j,q_t;p_s)}~~~\Label{Pen-coeff-final}\eea
but the definition of $\b_s^{(q_i,q_j,q_t;p_s)}$ now includes mass parameters:
\bea & & \b_s^{(q_i,q_j,q_t;p_s)} \equiv
\b_s^{(K_i,K_j,K_t;P_s)}~~~\Label{bs-k=3}\\ & =&
-{(K_i^2+M_1^2-m_i^2) \eps(P_s,K_j,K, K_t)+(K_j^2+M_1^2-m_j^2)
\eps(K_i,P_s,K, K_t)\over K^2 \eps(K_i,K_j,K, K_t)}\nonumber \\ & &
- {(K^2+M_1^2-M_2^2) \eps(K_i,K_j,P_s, K_t)+(K_t^2+M_1^2-m_t^2)
\eps(K_i,K_j,K, P_s)\over K^2 \eps(K_i,K_j,K, K_t)}\nonumber \eea
The expression (\ref{bs-k=3}) is symmetric in $K_i,K_j,K_t,K$
(recall that $M_2$ is the mass associated with $K$ in this context).

The box coefficients are given by
\bea C[Q_i,Q_j]_{k=3} & = & {(K^2)^{2+n}\over 2} \left(
{\prod_{s=1}^{n+3} \gb{P_{ji;1}|{\W R}_s(u)|P_{ji;2}}\over \gb{P_{ji;1}|K|P_{ji;2}}^{n+2}\gb{P_{ji;1}|{\W Q}_t(u)|P_{ji;2}}}\right.
\nonumber \\
& & \left.- <\prod_{s=1}^{n+3} \b_s^{(q_i,q_j,q_t;p_s)}{\gb{P_{ji;1}|K|P_{ji;2}}\over \gb{P_{ji;1}|{\W
Q}_t(u)|P_{ji;2}}}+\{ P_{ji;1}\leftrightarrow
P_{ji;2}\}\right).~~\Label{k=3-final-1} \eea
The derivation of (\ref{k=3-final-1}) involved the result from the
case $k=2$.  All mass-dependence is already included in the
definitions (\ref{def-alpha-box-k2}) and (\ref{W-R-S}), along with the
similarly defined vector $Q_t(u)$:
\bea  {\W R}_s(u) &=& {p_s\cdot q_0^{(q_i,q_j,K)}\over
(q_0^{(q_i,q_j,K)})^2}
(\a^{(q_i,q_j)}(u)-1)(-\b q_0^{(q_i,q_j,K)}) + R_s(u=0)~~~\Label{W-R-S-1}\\
\W Q_t(u) &= &{q_t\cdot q_0^{(q_i,q_j,K)}\over
(q_0^{(q_i,q_j,K)})^2} (\a^{(q_i,q_j)}(u)-1)(-\b
q_0^{(q_i,q_j,K)})+Q_t(u=0)~~~\Label{W-Q-t-1}\eea

\subsubsection{The case $k\geq 4$}

In the derivation of the formulas, we introduce the following
functions:
\bea \gamma_s^{(K_i,K_j,K_s,K_t)} & = & {(K_i^2+M_1^2-m_i^2)
\eps(K,K_j,K_s, K_t)+(K_j^2+M_1^2-m_j^2) \eps(K_i,K,K_s, K_t)\over
K^2 \eps(K_i,K_j,K, K_t)}\nonumber \\ & & + {(K_s^2+M_1^2-m_s^2)
\eps(K_i,K_j,K, K_t)+(K_t^2+M_1^2-m_t^2) \eps(K_i,K_j,K_s,K)\over
K^2 \eps(K_i,K_j,K, K_t)}\nonumber
\\ & & -{\eps(K_i,K_j,K_s, K_t)\over \eps(K_i,K_j,K,
K_t)}~~~~\Label{k4-gamma}\eea
The numerator of $\gamma_s^{(K_i,K_j,K_s,K_t)}$ is symmetric in
$K_i,K_j,K_s, K_t$; the denominator breaks this symmetry by singling
out $K_s$.

We find the following results.

The pentagon coefficients are given by
\bea C[Q_i, Q_j, Q_t] = (K^2)^{3+n}
{\prod_{s=1}^{n+3}
\b_s^{(q_i,q_j,q_t;p_s)} \over
\prod_{w=1,w\neq i,j,t}^k \gamma_w^{(K_i,K_j,K_w,K_t)}}.
\eea

The box coefficients are given by
\bea C[Q_i,Q_j]_{k\geq 4} & = & {(K^2)^{2+n}\over
2}\left\{{\prod_{s=1}^{k+n} \gb{P_{ji;1}| {\W R}_s(u)
|P_{ji;2}}\over \gb{P_{ji;1}|K |P_{ji;2}}^{n+2}
\prod_{t=1,t\neq
i,j}^k\gb{P_{ji;1}|\W Q_t(u) |P_{ji;2}}}\right.\nonumber  \\
& & \left.-\sum_{t=1,t\neq i,j}^k
{ \prod_{s=1}^{n+k}
\b_s^{(q_i,q_j,q_t;p_s)} \over \prod_{w=1,w\neq i,j,t}^k
\gamma_w^{(K_i,K_j;K_w,K_t)} }
{
\gb{P_{ji;1}|K|P_{ji;2}}\over \gb{P_{ji;1}|\W
Q_t(u)|P_{ji;2}} }\right\}\nonumber \\ & & +\{
P_{ji;1}\leftrightarrow
P_{ji;2}\}~~~\Label{k=4-box-equiv}\eea
Again, all the $u$-dependence is concentrated in $\W R(u)$ and $\W Q(u)$.
The definitions of $P_{ji;a}$, $\W R_s(u)$, $\W Q_t(u)$,
$\b_s^{(q_i,q_j,q_t;p_s)}$, and $\gamma_w^{(K_i,K_j;K_w,K_t)}$ are given in (\ref{pjia}),  (\ref{W-R-S-1}),
(\ref{W-Q-t-1}),  (\ref{bs-k=3}), and (\ref{k4-gamma}),  respectively.




\begin{thebibliography}{999}

\bibitem{Bern:1994zx}
  Z.~Bern, L.~J.~Dixon, D.~C.~Dunbar and D.~A.~Kosower,
  Nucl.\ Phys.\ B {\bf 425}, 217 (1994)
  [arXiv:hep-ph/9403226].

\bibitem{Bern:1994cg}
  Z.~Bern, L.~J.~Dixon, D.~C.~Dunbar and D.~A.~Kosower,
  Nucl.\ Phys.\  B {\bf 435}, 59 (1995)
  [arXiv:hep-ph/9409265].

\bibitem{MasterIntegrals}{
  G.~'t Hooft and M.~J.~G.~Veltman,
  Nucl.\ Phys.\  B {\bf 153}, 365 (1979); \\
Z.~Bern, L.~J.~Dixon and D.~A.~Kosower,
Phys.\ Lett.\  B {\bf 302}, 299 (1993)
[Erratum-ibid.\  B {\bf 318}, 649 (1993)]
[hep-ph/9212308];\\
Z.~Bern, L.~J.~Dixon and D.~A.~Kosower,
Nucl.\ Phys.\  B {\bf 412}, 751 (1994)
[hep-ph/9306240];\\
J.~Fleischer, F.~Jegerlehner and O.~V.~Tarasov,
Nucl.\ Phys.\  B {\bf 566}, 423 (2000)
[hep-ph/9907327];\\
T.~Binoth, J.~P.~Guillet and G.~Heinrich,
Nucl.\ Phys.\  B {\bf 572}, 361 (2000)
[hep-ph/9911342];\\
G.~Duplan\v{c}i\'c and B.~Ni\v{z}i\'c,
Eur.\ Phys.\ J.\  C {\bf 35}, 105 (2004)
[hep-ph/0303184]; \\
  R.~K.~Ellis and G.~Zanderighi,
  arXiv:0712.1851 [hep-ph].
}


\bibitem{Cachazo:2004by}
  F.~Cachazo, P.~Svrcek and E.~Witten,
  JHEP {\bf 0410}, 077 (2004)
  [arXiv:hep-th/0409245].

\bibitem{Bena:2004xu}
  I.~Bena, Z.~Bern, D.~A.~Kosower and R.~Roiban,
  Phys.\ Rev.\  D {\bf 71}, 106010 (2005)
  [arXiv:hep-th/0410054].


\bibitem{Cachazo:2004dr}
  F.~Cachazo,
  arXiv:hep-th/0410077.

\bibitem{Britto:2004nj}
  R.~Britto, F.~Cachazo and B.~Feng,
  Phys.\ Rev.\  D {\bf 71}, 025012 (2005)
  [arXiv:hep-th/0410179].

\bibitem{Britto:2004nc}
  R.~Britto, F.~Cachazo and B.~Feng,
  Nucl.\ Phys.\ B {\bf 725}, 275 (2005)
  [arXiv:hep-th/0412103].

\bibitem{Britto:2005ha}
  R.~Britto, E.~Buchbinder, F.~Cachazo and B.~Feng,
  Phys.\ Rev.\ D {\bf 72}, 065012 (2005)
  [arXiv:hep-ph/0503132].

\bibitem{Brandhuber:2005jw}
  A.~Brandhuber, S.~McNamara, B.~J.~Spence and G.~Travaglini,
  JHEP {\bf 0510}, 011 (2005)
  [arXiv:hep-th/0506068].

\bibitem{Britto:2006sj}
  R.~Britto, B.~Feng and P.~Mastrolia,
  Phys.\ Rev.\ D {\bf 73}, 105004 (2006)
  [arXiv:hep-ph/0602178].


\bibitem{Anastasiou:2006jv}
  C.~Anastasiou, R.~Britto, B.~Feng, Z.~Kunszt and P.~Mastrolia,
  Phys.\ Lett.\  B {\bf 645}, 213 (2007)
  [arXiv:hep-ph/0609191].

\bibitem{Mastrolia:2006ki}
  P.~Mastrolia,
  Phys.\ Lett.\  B {\bf 644}, 272 (2007)
  [arXiv:hep-th/0611091].

\bibitem{Britto:2006fc}
  R.~Britto and B.~Feng,
  Phys.\ Rev.\  D {\bf 75}, 105006 (2007)
  [arXiv:hep-ph/0612089].



\bibitem{Anastasiou:2006gt}
  C.~Anastasiou, R.~Britto, B.~Feng, Z.~Kunszt and P.~Mastrolia,
  JHEP {\bf 0703}, 111 (2007)
  [arXiv:hep-ph/0612277].




\bibitem{Britto:2007tt}
  R.~Britto and B.~Feng,
  arXiv:0711.4284 [hep-ph].





\bibitem{OPP}{
  G.~Ossola, C.~G.~Papadopoulos and R.~Pittau,
  Nucl.\ Phys.\  B {\bf 763}, 147 (2007)
  [arXiv:hep-ph/0609007];

  JHEP {\bf 0707}, 085 (2007)
  [arXiv:0704.1271 [hep-ph]];

  arXiv:0711.3596 [hep-ph];

  arXiv:0802.1876 [hep-ph];
}

\bibitem{Forde:2007mi}
  D.~Forde,
  Phys.\ Rev.\  D {\bf 75}, 125019 (2007)
  [arXiv:0704.1835 [hep-ph]].

\bibitem{Kilgore:2007qr}
  W.~B.~Kilgore,
  arXiv:0711.5015 [hep-ph].

\bibitem{BjerrumBohr:2007vu}
  N.~E.~J.~Bjerrum-Bohr, D.~C.~Dunbar and W.~B.~Perkins,
  arXiv:0709.2086 [hep-ph].

\bibitem{Ellis:2007br}
  R.~K.~Ellis, W.~T.~Giele and Z.~Kunszt,
  arXiv:0708.2398 [hep-ph].

\bibitem{Giele:2008ve}
  W.~T.~Giele, Z.~Kunszt and K.~Melnikov,
  arXiv:0801.2237 [hep-ph].




\bibitem{Cachazo:2004kj}
  F.~Cachazo, P.~Svrcek and E.~Witten,
  JHEP {\bf 0409}, 006 (2004)
  [arXiv:hep-th/0403047].

\bibitem{SpinorFormalism}{
  F.~A.~Berends, R.~Kleiss, P.~De Causmaecker, R.~Gastmans and T.~T.~Wu,
  Phys.\ Lett.\  B {\bf 103}, 124 (1981).

  P.~De Causmaecker, R.~Gastmans, W.~Troost and T.~T.~Wu,
  Nucl.\ Phys.\  B {\bf 206}, 53 (1982).

  R.~Kleiss and W.~J.~Stirling,
  Nucl.\ Phys.\  B {\bf 262}, 235 (1985).


  R.~Gastmans and T.~T.~Wu,
{\it  Oxford, UK: Clarendon (1990) 648 p. (International series of
  monographs on physics, 80)}

  Z.~Xu, D.~H.~Zhang and L.~Chang,
  Nucl.\ Phys.\  B {\bf 291}, 392 (1987).

  J.~F.~Gunion and Z.~Kunszt,
  Phys.\ Lett.\  B {\bf 161}, 333 (1985).
}



\bibitem{DDimU}{
Z.~Bern, L.~J.~Dixon, D.~C.~Dunbar and D.~A.~Kosower,
Phys.\ Lett.\ B {\bf 394}, 105 (1997) [hep-th/9611127].

W.~L.~van Neerven,
Nucl.\ Phys.\  B {\bf 268}, 453 (1986).

A.~Brandhuber, S.~McNamara, B.~J.~Spence and G.~Travaglini,
JHEP {\bf 0510}, 011 (2005) [hep-th/0506068].
}

\bibitem{Bern:1995db}
  Z.~Bern and A.~G.~Morgan,
  Nucl.\ Phys.\  B {\bf 467}, 479 (1996)
  [arXiv:hep-ph/9511336].

\bibitem{Rozowsky:1997dm}
  J.~S.~Rozowsky,
  arXiv:hep-ph/9709423.



\bibitem{Maitre:2007jq}
  D.~Maitre and P.~Mastrolia,
  arXiv:0710.5559 [hep-ph].

\bibitem{Ellis:1987xu}
  R.~K.~Ellis, I.~Hinchliffe, M.~Soldate and J.~J.~van der Bij,
  Nucl.\ Phys.\  B {\bf 297}, 221 (1988).


\bibitem{bfy}
R.~Britto, B.~Feng and G.~Yang, in preparation.

\end{thebibliography}
\end{document}